\documentclass[12pt]{iopart}

\usepackage{graphicx}
\usepackage{dcolumn}
\usepackage{color}
\usepackage{bm}
\usepackage{iopams}
\pdfoutput=1
\RequirePackage[colorlinks,citecolor=blue,urlcolor=magenta,linkcolor=blue]{hyperref}
\input epsf

\begin{document}
\title[Equilibrium Configuration of Perfect Fluid]{Equilibrium configuration of perfect fluid orbiting around 
black holes in some classes of alternative gravity theories}
\author{Sumanta Chakraborty}
\address{IUCAA, Post Bag 4, Ganeshkhind,
Pune University Campus, Pune 411 007, India}
\ead{sumantac.physics@gmail.com;~~sumanta@iucaa.ernet.in}
\date{\today}
\begin{abstract}
The hydrodynamic behaviour of perfect fluid orbiting around black holes in spherically symmetric spacetime for various alternative gravity theories has been investigated. For this purpose we have assumed an uniform distribution for the angular momentum density of the rotating perfect fluid. The contours of equipotential surfaces are illustrated in order to obtain the nature of inflow and outflow of matters. It has been noticed that, the marginally stable circular orbits originating from decreasing angular momentum density lead to closed equipotential surfaces along with cusp allowing existence of accretion disks. On the other hand, the growing part of angular momentum density exhibits central rings for which stable configurations are possible. However inflow of matter is prohibited. Among the solutions discussed in this work, the charged $F(R)$ gravity and Einstein-Maxwell-Gauss-Bonnet solution exhibit inflow and outflow of matter with central rings present. These varied accretion disk structure of perfect fluid attribute these spacetimes astrophysical importance. The effect of higher curvature terms predominantly arises from region near the black hole horizon. Hence the structural difference of accretion disk in modified gravity theories with comparison to general relativity may act as an experimental probe for these alternative gravity theories. 
\end{abstract}
\maketitle
\section{Introduction}\label{equiintro}

General Relativity (GR) is a very successful theory and is the best contender, so far to describe the geometrical properties of the spacetime. It has passed through all the experimental and observational tests so far, ranging from local gravity tests like perihelion precession and bending of light to precision tests using pulsars \cite{Kramer2014,Kramer2008,Kramer2009,Peterson97,Osterbrock91}. In spite of these outstanding success for GR there are still unresolved issues. These include: the problem of Dark energy and the problem of inflation. Phrased in a different way, the reason for accelerated expansion of the universe at both very small scale and very large scale is unknown \cite{Riess98,Perlmutter99,Nojiri07}. These issues are generally dealt with by assuming existence of additional matter component in the universe, like cosmological constant, fluid with complicated equation of state, scalar fields etc. However all these models are plagued with several issues, e.g., coupling with usual matter, consistency with elementary particle theories and consistency of formulation \cite{Felice10,Faraoni07,Nojiri2011}. These results prompted research in a new direction by modifying the Einstein-Hilbert (EH) action for GR itself. 

This idea of modifying the EH action stem from the belief that GR is just a low energy approximation of some underlying fundamental theory \cite{Buchbinder1992,Vassilevich2003}. In this spirit the classical generalizations of EH action should explain both early-time inflation and late-time cosmic acceleration without ever introducing additional matter components in the energy momentum tensor. There exist large number of ways in which the standard EH action can be modified by introducing non-linear terms. However the criteria that the field equations should remain second order in the dynamical variable (otherwise some ghost fields would appear) uniquely fixes the action to be the Lanczos-Lovelock action \cite{Lanczos1932,Lovelock1971,Padmanabhan2010b}. The Lanczos-Lovelock Lagrangian has this special property that though the Lagrangian contains higher curvature terms the field equation turns out to be of second order \cite{Padmanabhan13,Padmanabhan2011b}. One of the important model is the second order correction term in the gravity action in addition to EH term, called Gauss-Bonnet (GB) Lagrangian.

Another such model explaining the above mentioned problems is obtained by replacing $R$, the scalar curvature in the EH action by some arbitrary function of the scalar curvature $F(R)$. This alternative theory for describing gravitational interaction is very interesting in its on right, for it can provide an explanation to a large number of experimental phenomenon. These include: late time cosmic acceleration \cite{Bamba08}, early power law inflation \cite{Starobinsky80}, problem of singularity in strong gravity regime \cite{Abdalla05} along with possible detection of gravitational waves \cite{Corda08}. Compatibility with Newtonian and post-Newtonian approximations as well as estimations of cosmological parameters are shown in \cite{Nojiri07,Takada04}. $F(R)$ theories can also bypass a long known instability problem, called `Ostrogradski' instability \cite{Woodard07} and is capable of explaining recent graviton mass bound in LHC, with possible constraints on the model itself \cite{Chakraborty2014}.

There also exist quiet a few gravity models, which comes into existence from some low energy string inspired theories. For example the GB Lagrangian discussed earlier is actually a low energy realization of a string inspired model \cite{Boulware1985}.  In this context, we should mention that there exist a natural generalization of the Reissner-Nordstr\"om (RN) solution in GR to that in string inspired models with dilaton coupling. The dilaton field couples with the electromagnetic field tensor $F_{\mu \nu}$ in such a fashion that every solution with non-zero $F_{\mu \nu}$ couples with the dilaton field \cite{Garfinkle91,Coleman83,Vega88,Witten62}. 

In this work our main aim is to obtain the equilibrium configuration of perfect fluid rotating around black holes in these alternative gravity theories. This in turn will allow us to address the new features that come out in comparison to their general relativistic counterpart. For this purpose we have first studied the equilibrium  configuration of fluid in a general static spherically symmetric spacetime. Having derived various quantities of interest in this general context, we apply them to the black hole solutions in various alternative gravity theories. This can be achieved by substituting the metric elements describing spherically symmetric spacetime in these alternative theories into our general result. Throughout this work we have mainly concentrated on three alternative theories, the $F(R)$ gravity theory, dilaton induced gravity theory and finally the Einstein-Maxwell-Gauss-Bonnet (EMGB) gravity.

After deriving the results for a general static spherically symmetric spacetime, as a warm up we have applied our results to the RN solution in Einstein gravity. Then we have applied these results to a charged black hole solution in $F(R)$ gravity theory. It turns out that the equipotential contours of the perfect fluid rotating around black hole in this gravity model are quiet different from its GR counterpart. Identical features are shared by the equilibrium configuration of perfect fluid rotating around black hole in the EMGB theory as well. However the equipotential contours of perfect fluid rotating around black holes in dilaton gravity are different from the previous two situations. Still the equipotential contours exhibit features distinct in comparison to the respective GR solution. All these results suggest that the equilibrium configurations of perfect fluid rotating around black holes in alternative gravity theories have structures different from GR. By comparing these alternative theories under one roof, as presented in this work, we have been able to understand the structure of accretion disks and in-fall of matter to them in some detail. This might help to understand the accretion disk structure around other black hole solutions by direct comparison with results presented in this work. 

Since the accretion disk is intimately connected with various astrophysical processes, it is natural to ask for some observational consequences of our result. We should stress at this point that though the accretion disk structure gets modified with outflow and inflow of matter associated due to introduction of alternative theories it is unlikely to be observed in an observational test. These results have to do with the fact that the accretion disk structure across a black hole in these alternative theories are determined by the dimensionless parameter $y=\Lambda M^{2}/12$ in geometrized units, where $\Lambda$ is the cosmological constant and $M$ being the black hole mass. In order to have stable circular orbits necessary for existence of accretion disc, the cosmological parameter has a restriction $y<y_{e}\sim 4.8 \times 10^{-5}$ as we have shown later. Along with we can use cosmological tests using magnitude-redshift relation for supernova with the measurements of Cosmic Microwave Background (CMB) fluctuation \cite{Spergel2003,Spergel2007} implying $\Lambda \sim 10^{-56}~\textrm{cm}^{-2}$. This leads to very low values of $y$ for realistic black holes. For example, in case of an extremely massive black hole as seen in quasar TON 618, with mass $\sim 6.6 \times 10^{10}M_{\bigodot}$ \cite{Shemmer2004} leads to $y\sim 4.1\times 10^{-25}$. This suggests that the parameter space for observations in astrophysical scenarios has the value of cosmological parameter $y\sim 10^{-30}$. This is quiet small compared to the value $y\sim 10^{-7}$ presented in the text for being of astrophysical importance  \cite{Abramowicz97,Mirabel99,Fender99,Novikov73,Kozlowski78,Abramowicz78,Jaroszynski80,Lynden69,Blandford87}. Thus for normal astrophysical systems it is difficult to detect any departure from the result predicted by GR. Even for quiet massive systems departure from GR, with observed accuracy has not been found to date \footnote{one such system where the effect can be observed is a pulsar-White Dwarf (WD) system. Such a recent system, PSR J0348+0432 has a huge mass for the Neutron star, making sensitive test for the strong gravity regime possible. To date the orbital period decay of this system is in agreement with GR \cite{Antoniadis2013}. Another such system PSR J1738+0333 also shows results consistent with GR \cite{Freire2012}.}. Such small values are out of the parameter space for present day astrophysical observations, making the test for these alternative theories difficult. 

However there is another window to look for signatures of these alternative theories. For primordial black holes in the early universe the \emph{effective} cosmological constant was higher \cite{Stuchlik2005} and consequently $y$ would had a much larger value of comparable to $y_{e}$. Thus in the early universe the accretion disk structure around primordial black holes may posses the features discussed in this work. But this is also far from being observed experimentally. In spite of these difficulties in relating to observational results, the work stands on its own, for it describes the accretion structure in alternative theories and its departure from the GR results. For some choice of the parameter space the structure in alternative theories are much rich and differ significantly from that in GR. This analysis brings out the fact that introduction of higher curvature terms in the EH action alters the accretion disk structure of perfect fluid orbiting the black hole in a non-trivial manner with more structures included.

The paper is organized as follows: In section \ref{eqfluid} we have illustrated the Boyer's condition \cite{Boyer65,Abramowicz74,Stuchlik00} in order to have an equilibrium configuration of the perfect fluid. Then in the next section \ref{equigen} we have derived various physical quantities for a general static spherically symmetric spacetime, which we have subsequently applied to determine equipotential surfaces of fluid moving around various black hole solutions in different gravity theories. Among these alternative theories we have discussed the RN solution, topologically charged black hole in $F(R)$ theory, charged black hole in dilaton gravity and finally black hole in EMGB theory in \ref{equialt}. At the end we conclude with a discussion on our results.

\section{Equilibrium configuration of rotating perfect fluid}\label{eqfluid}

In this section we briefly summarize the well known results
regarding general theory of equipotential surfaces inside any
relativistic, differentially rotating, perfect fluid body \cite{Boyer65,Abramowicz74}. 
This has also been applied to configurations of perfect fluid rotating in the stationary
and axi-symmetric spacetimes \cite{Kozlowski78,Abramowicz78,Jaroszynski80}. 
In standard coordinate
system the spacetime is described by the following line element,
\begin{equation}\label{eq1}
ds^{2}=g_{tt}dt^{2}+2g_{t\phi}dtd\phi + g_{\phi \phi}d\phi ^{2}+g_{rr}dr^{2}+g_{\theta \theta}d\theta ^{2}
\end{equation}
where the metric elements depend neither on time coordinate $t$ nor on azimuthal coordinate $\phi$.
Hence energy and angular momenta are two conserved quantities
in this spacetime implying existence of two
killing vectors $\frac{\partial}{\partial t}$ and $\frac{\partial}{\partial \phi}$.
We shall consider perfect fluid rotating in the $\phi$ direction.
Then its four velocity field $U^{\mu}$ has only two non-zero components,
\begin{equation}\label{eq2}
U^{\mu}=\left(U^{t},U^{\phi},0,0\right)
\end{equation}
which can be functions of the coordinates $(r,\theta)$. The stress
energy tensor of the perfect fluid is,
\begin{equation}\label{eq3}
T^{\mu}_{\nu}=(p+\epsilon)U^{\mu}U_{\nu}+p\delta ^{\mu}_{\nu}
\end{equation}
where $\epsilon$ and $p$ denote the total energy density and the
pressure of the fluid. The rotating fluid can be characterized by
the vector fields of the angular velocity $\Omega$, and the
angular momentum per unit mass (angular momentum density) $\ell$,
defined by
\begin{equation}\label{eq4}
\Omega = \frac{U^{\phi}}{U^{t}};\qquad  \ell =-\frac{U_{\phi}}{U_{t}}
\end{equation}
These vector fields are related to the metric elements by the following result
\begin{equation}\label{eq5}
\Omega = -\frac{g_{t\phi}+\ell g_{tt}}{g_{\phi \phi}+\ell g_{t\phi}}
\end{equation}
In static spacetimes ($g_{t\phi}=0$), the above relation reduces to the simple formula
\begin{equation}\label{eq6}
\frac{\Omega}{\ell}=-\frac{g_{tt}}{g_{\phi \phi}}
\end{equation}
The surfaces of constant $\ell$ and $\Omega$ are called von Zeipel's cylinders.
These surfaces do not depend on the rotation law for fluids
in static spacetimes but do depend on the rotation law for
stationary spacetimes \cite{Kozlowski78}.

Projecting the stress energy tensor conservation law $\nabla _{\nu}T^{\mu \nu}=0$
onto the hypersurface
orthogonal to the four velocity $U^{\mu}$ by the projective tensor
$h_{\mu \nu}=g_{\mu \nu}+U_{\mu}U_{\nu}$, we obtain the relativistic Euler equation in the form,
\begin{equation}\label{eq7}
\frac{\partial _{\mu}p}{p+\epsilon}=-\partial _{\mu}(\ln U_{t})+\frac{\Omega \partial _{\mu}\ell}{1-\Omega \ell}
\end{equation}
where
\begin{equation}\label{eq8}
(U_{t})^{2}=\frac{g_{t\phi}^{2}-g_{tt}g_{\phi \phi}}{g_{\phi \phi}+2\ell g_{t\phi}+\ell ^{2}g_{tt}}
\end{equation}
The solution to the relativistic Euler equation given in Eq. (\ref{eq7}) can be obtained by defining the potential $W(r,\theta)$  whose ``equipotential surfaces" are surfaces of constant pressure through the following relation \cite{Abramowicz78,Stuchlik00,Stuchlik2009}:
\begin{equation}\label{eq9}
\int _{0}^{p}\frac{dp}{p+\epsilon} = W_{in}-W
\end{equation}
which leads to
\begin{equation}\label{eq10}
W_{in}-W= ln(U_{t})_{in} - ln(U_{t})+\int _{\ell _{in}}^{\ell} \frac{\Omega d\ell}{1-\Omega \ell}
\end{equation}
The subscript ``in" in the above equations refer to the inner edge of the disk. For an alternative definition and thus derivation  of Boyer's conditions see Ref. \cite{Abramowicz78,Fishbone76,Fishbone77}.
The equipotential surfaces are determined by the condition,
$W(r,\theta)=\textrm{constant}$ and
in a given spacetime $W$ can be found from Eq. ($\ref{eq10}$), if the rotation
law $\Omega = \Omega (\ell)$ is known. The surfaces of constant pressure are being determined
by Eq. ($\ref{eq9}$). The structure of thick accretion disks can also be found using
the accurate Newtonian framework called Pseudo-Newtonian Potential method \cite{Paczynski80,Abramowicz80}.

\section{Equilibrium configuration of a Perfect Fluid for a general spherically symmetric spacetime}\label{equigen}

In this section we shall start with a general metric ansatz. This is very appropriate for
describing various static spherically symmetric solutions in both Einstein gravity or other
alternative gravity theories \cite{Chakraborty11,Chakraborty13}.
The metric ansatz is simple enough that
it can be easily generalized to arbitrary number of spacetime dimensions. The general
metric ansatz is taken as,
\begin{equation}\label{geneq1}
ds^{2}=-f(r)dt^{2}+f(r)^{-1}dr^{2}+r^{2}d\Omega ^{2}
\end{equation}
Note that the characteristic property of equipotential surfaces
is retained in this simplest choice, which is uniform
distribution of angular momentum density $\ell$ \cite{Jaroszynski80}. 
The importance of the above statement enhances
considerably from the fact that marginally stable
configurations comes into existence under the condition \cite{Seguin75} 
\begin{equation}\label{geneq1a}
\ell (r,\theta)=\textrm{const}
\end{equation}
This holds for any rotating fluid irrespective of its rotation law for static spacetime (however for stationary spacetime the above condition depends on the rotation law \cite{Kozlowski78}). 
Using which we can obtain the following
expression for equipotential surfaces from Eq. (\ref{eq10}) as
\begin{equation}\label{geneq1b}
W(r,\theta)=\ln U_{t}(r,\theta)
\end{equation}
Note that the quantity $U_{t}(r,\theta)$ depends only on the metric and is being completely
determined by Eq. (\ref{geneq1a}). The equation for these equipotential surfaces
are to be given by the following relation $\theta =\theta (r)$, which can be
obtained by solving the following differential equation
\begin{equation}\label{geneq1c}
\frac{d\theta}{dr}=-\frac{\partial p/\partial r}{\partial p/\partial \theta}
\end{equation}
For the $\ell =\textrm{constant}$ configurations the above expression leads to,
\begin{equation}\label{geneq1d}
\frac{d\theta}{dr}=-\frac{\partial U_{t}/\partial r}{\partial U_{t}/\partial \theta}
\end{equation}
Having obtained all these basic tools we now proceed to determine
the equilibrium configuration of perfect fluid orbiting around a
black hole in a spacetime with a metric ansatz given by Eq.
(\ref{geneq1}). Thus the potential turns out to be
\begin{equation}\label{geneq2}
W(r,\theta)=\ln \left[\frac{r\sin \theta \sqrt{f(r)}}{\sqrt{r^{2}\sin ^{2}\theta -\ell ^{2}f(r)}} \right]
\end{equation}
as well as the differential equation satisfied by $\theta$ leads to
\begin{equation}\label{geneq3}
\frac{d\theta}{dr}=\frac{r^{3}\sin ^{2}\theta f'(r)-2\ell ^{2}f(r)^{2}}{2r\ell ^{2}f(r)^{2}}\tan \theta
\end{equation}
where `prime' denotes derivative with respect to the radial
co-ordinate $r$. The insight about the $\ell =\textrm{constant}$
surfaces are gained by examination of the potential $W(r,\theta)$
in the equatorial plane $\theta =\pi /2$. From physical principles
the potential should be real and that eventually leads to the
following conditions
\begin{eqnarray}\label{geneq3a}
f(r)\geq 0
\\
r^{2} -\ell ^{2}f(r)\geq 0
\end{eqnarray}
The first condition would lead to static regions outside the black
hole horizon where the perfect fluid orbiting the black hole can
have equilibrium configurations. The second condition
can be expressed in an elegant form as:
\begin{equation}\label{geneq3b}
\ell ^{2}\leq \ell _{ph}^{2}\equiv \frac{r^{2}}{f(r)}
\end{equation}
where the function defined as $\ell _{ph}^{2}$ is the effective potential of photon geodesic,
with impact parameter $\ell =U_{\phi}/U_{t}$ \cite{Stuchlik99}. Note that the extremization
of the potential $W(r,\theta =\pi /2)$ being identical with the condition that pressure gradient
should vanish i.e. $\partial U_{t}/\partial r=0$ and $\partial U_{t}/\partial \theta =0$.
However since we are concerned with the equatorial plane ($\theta =\pi /2$), the criteria
$(\partial W/\partial r)=0$ is sufficient for maximization of $W(r,\theta =\pi/2)$ leading to
\begin{equation}\label{geneq3c}
\frac{\partial U_{t}(r,\theta =\pi /2)}{\partial r}=\frac{r^{3}f'(r)-2\ell ^{2}f(r)^{2}}{2\sqrt{f}\left[r^{2}-\ell ^{2}f(r)\right]^{3/2}}
\end{equation}
It is well known that the extrema of the potential on the
equatorial plane $W(r,\theta =\pi/2)$ corresponds to motion along
circular geodesics. Thus the angular momentum distribution $\ell
_{K}^{2}$ obtained from the condition $\partial _{r}U_{t}=0$ represents the angular momentum density for circular orbits. From Eq. (\ref{geneq3c}) the following estimation of angular momentum density $\ell _{K}^{2}$ becomes possible
\begin{equation}\label{geneq4}
\ell _{K}^{2}=\frac{r^{3}f'(r)}{2f(r)^{2}}
\end{equation}
With this identification of $\ell _{K}^{2}$, the extrema for the potential $W(r,\theta =\pi/2)$ can be obtained as
\begin{equation}\label{geneq4a}
W_{extrema}(r,\theta =\pi /2)=\ln E_{c}
\end{equation}
which determines
\begin{equation}\label{geneq5}
E_{c}(r)=\frac{\sqrt{2}f(r)}{\sqrt{2f(r)-rf'(r)}}
\end{equation}
to be the specific energy attributed for motion along circular
geodesics. Thus in order to summarize our results, we have
discussed that one of the most important properties of the
potential $W(r,\theta)$ is that its behaviour at equatorial plane
$\theta =\pi /2$ completely determines all the parameters under
interest through $\ell _{ph}^{2}$ and $\ell _{K}^{2}$. From
these two angular momentum densities we will be able to introduce more parameters in order to
explain the behaviour of the fluid configuration in an compact and
understandable fashion.

For that purpose there exists two equations of utmost importance. One of them comes from
the fact that minima of $\ell _{ph}^{2}$ gives the photon radius $r_{ph}$ and is determined
by solving the equation,
\begin{equation}\label{geneq5a}
rf'(r)=2f(r)
\end{equation}
Also note that $\ell _{K}^{2}$ diverges at the horizon given by $f(r)=0$. Since we have,
\begin{equation}\label{geneq6}
\frac{\partial \ell _{K}^{2}}{\partial r}=\frac{2f(r)^{2}\left[3r^{2}f'(r)+r^{3}f''(r)\right]-4r^{3}f(r)f'(r)^{2}}{4f(r)^{4}}
\end{equation}
the local extrema of the angular momentum density $\ell _{K}^{2}$ is being determined by the following equation,
\begin{equation}\label{geneq7}
4r^{3}f'(r)^{2}=2f(r)\left[3r^{2}f'(r)+r^{3}f''(r) \right]
\end{equation}
Hence all the relevant quantities necessary to uniquely characterize the fluid motion are
derived for a general metric ansatz. We will henceforth use these results frequently
throughout the later part of this work.

\section{Equipotential Surfaces of Marginally Stable Configurations Orbiting Around Black Holes in Various Gravity Theories }
\label{equialt}

In this section we will apply the results derived in the previous section to spherically symmetric and static spacetimes in both GR and alternative gravity theories. We will start with the Reissner-Nordstr\"om solution in general relativity and then subsequently will generalize to alternative gravity theories. Among these alternative theories, we will consider the black hole solution in $F(R)$ gravity first, then Einstein-Maxwell gravity with dilaton field, finally specializing to Einstein-Maxwell-Gauss-Bonnet gravity (similar aspects within GR framework has been considered in \cite{Kovar2014,Slany2013}).

\subsection{Reissner-Nordstr\"om Black Hole}\label{equiRN}

Before proceeding to alternative theories we shall first consider a solution in GR itself. This will help the reader to understand various physical quantities of interest in the other theories better.
The line element for the RN black hole has the following expression
\begin{equation}\label{eqrn1}
ds^{2}=-\left(1-\frac{2M}{r}+\frac{Q^{2}}{r^{2}}\right)dt^{2}+\left(1-\frac{2M}{r}+\frac{Q^{2}}{r^{2}} \right)^{-1}dr^{2}+r^{2}d\Omega ^{2}
\end{equation}
where $Q$ is the charge of the black hole and $M$ is its mass. The potential $W(r,\theta)$, introduced by Eq. (\ref{eq9}) and Eq. (\ref{geneq2}) for the above metric takes the following form
\begin{equation}\label{eqrn2}
W(r,\theta)=\frac{1}{2}\ln \left[\frac{r^{2}\sin ^{2}\theta\left(r^{2}-2Mr+Q^{2}\right)}{r^{4}\sin ^{2}\theta-\left(r^{2}-2Mr+Q^{2}\right)\ell ^{2}} \right]
\end{equation}
Hence from the reality criteria of the potential we arrive at the following constraints
\begin{eqnarray}\label{eqrn3}
\left(1-\frac{2M}{r}+\frac{Q^{2}}{r^{2}}\right)\geq 0
\\
\ell ^{2}\leq \ell _{ph}^{2}=\frac{r^{4}}{r^{2}-2Mr+Q^{2}}
\label{eqrn3a}
\end{eqnarray}
where, $\ell _{ph}$ corresponds to angular momentum associated with the photon orbit. Also we have the following result for $(d\theta /dr)$ from Eq. (\ref{geneq3}) as
\begin{equation}\label{eqrn4}
\frac{d\theta}{dr}=\frac{r^{4}\left(Mr-Q^{2}\right)\sin^{2}\theta -\ell ^{2}\left(r^{2}-2Mr+Q^{2}\right)^{2}}
{r\ell ^{2}\left(r^{2}-2Mr+Q^{2}\right)^{2}}\tan \theta
\end{equation}
Now there are two horizons of this space time, one is the regular event horizon located at $r=r_{eh}$ and the other horizon, known as the Cauchy horizon located at $r=r_{ch}$, have the following expressions
\begin{eqnarray}
r_{eh}&=&M+\sqrt{M^{2}-Q^{2}}
\\
r_{ch}&=&M-\sqrt{M^{2}-Q^{2}}
\end{eqnarray}
Another important radius is the photon circular orbit radius, whose location in this spacetime is given by the following expression:
\begin{equation}\label{eqrn5}
r_{ph}=\frac{3M\pm \sqrt{9M^{2}-8Q^{2}}}{2}
\end{equation}
The extrema of the potential $W(r,\theta)$ on the equatorial plane is being determined by an angular momentum density $\ell _{K}^{2}$ from Eq. (\ref{geneq4}) with the following expression
\begin{equation}\label{eqrn5a}
\ell _{K}^{2}=\frac{r^{4}\left(Mr-Q^{2}\right)}{\left(r^{2}-2Mr+Q^{2}\right)^{2}}
\end{equation}
Then we also have the following expression for derivative of $\ell _{K}^{2}$:
\begin{equation}\label{eqrn6}
\frac{\partial \ell _{K}^{2}}{\partial r}=\frac{\left(r^{2}-2Mr+Q^{2}\right)\left(5Mr^{4}-4r^{3}Q^{2} \right)-4r^{4}\left(r-M\right)\left(rM-Q^{2}\right)}{\left(r^{2}-2r+Q^{2}\right)^{3}}
\end{equation}
The minima of the photon angular momentum can be obtained by plugging in Eq. (\ref{eqrn5}) into Eq. (\ref{eqrn3a}) which leads to:
\begin{equation}
\ell ^{2}_{ph(min)}=\frac{\left(3M+\sqrt{9M^{2}-8Q^{2}}\right)^{4}}
{8\left(3M^{2}-2Q^{2}+M\sqrt{9M^{2}-8Q^{2}}\right)}
\end{equation}
Thus we have presented all these results for a well known solution in GR to facilitate easy comparison with other gravity theories. This also brings out the key parameters involved in these computations which would be helpful as we go along discussing various alternative gravity theories.
\subsection{Charged Black Hole in $F(R)$ Gravity}\label{fluidfr}

In this section we will restrict ourselves to four dimensional spacetime. The action in this four dimensional spacetime with $F(R)$ gravity in presence of matter field can be presented as:
\begin{equation}\label{FRC01}
S=\frac{1}{16\pi}\int d^{4}x \sqrt{-g}\left[F(R)+\mathcal{L}_{matter} \right]
\end{equation}
In the above expression for action $F(R)$ is an arbitrary function of the Ricci scalar $R$ and $\mathcal{L}_{matter}$ is the Lagrangian for matter fields. Then variation of the above action with respect to the metric leads to the field equation \cite{Felice10,Smith03}:
\begin{equation}\label{FRC02}
R_{ab}F_{R}-\nabla _{a}\nabla _{b}F_{R}+\left(\square F_{R}-\frac{1}{2}F(R)\right)g_{ab}=T_{ab}^{matter}
\end{equation}
where $R_{ab}$ is the Ricci tensor, $F_{R}\equiv dF(R)/dR$ and $T_{ab}^{matter}$ is the standard matter stress-energy tensor, derived from the matter part of the Lagrangian $\mathcal{L}_{matter}$ as given in action (\ref{FRC01}). Also the trace of the above equation connects trace of stress-energy tensor to the scalar curvature. We are interested in determining spherically symmetric solution to the above field equation. Following the analysis presented in \cite{Hendi12} we arrive at a charged solution in this $F(R)$ gravity models with $F(R)=R-\lambda \exp(-\xi R)$. It should be noted as emphasized in Sec. \ref{equiintro}, that these modifications in Einstein-Hilbert action should pass all tests starting from clustering of galaxies down to solar system tests. For this special exponential correction factor it has been shown \cite{Cognola08} that there is no contradiction with solar system tests. Also the solutions of this model are indistinguishable from the standard Einstein-Hilbert solution except for a possible change in Newton's constant \cite{Zhang07}. The solution to the above field equation presented in Eq. (\ref{FRC02}) is topologically charged and can be presented following Eq. (\ref{geneq1}) as
\begin{equation}\label{FRC03}
f(r)=1-\frac{\Lambda}{3}r^{2}-\frac{M}{r}+\frac{Q^{2}}{r^{2}}
\end{equation}
where $\Lambda =\lambda\left[(4+2\xi R)/(8e^{\xi R})\right]$. In order for this metric ansatz to satisfy Eq. (\ref{FRC02}) we should have some constraint among the parameters appearing in our theory, which are the followings \cite{Hendi12}:
\begin{eqnarray}\label{FRC04}
1+\frac{\lambda \xi}{\exp(\xi R)}=0\\
\frac{\lambda}{\exp(\xi R)}+\frac{R}{2}\left(\frac{\lambda \xi}{\exp(\xi R)}-1\right)=0
\end{eqnarray}
This solution though looks similar to the charged black hole, its higher dimensional behaviour is different. The charge term in the above expression goes as $r^{-(d-2)}$ in $d$ dimension, while it goes as $r^{-2(d-3)}$ in standard charged higher dimensional black hole spacetime. Therefore it is interesting to ask whether this term is originating from the scalar-tensor representation of $F(R)$ gravity. In order to obtain the representation the standard way is to start with conformal transformations on the metric elements. However in this case no new physical interpretation can be obtained except its relation to the scale of the problem \cite{Faraoni07,Hendi12,Capone09}. 

The above result can also be interpreted in a slightly different manner, uncharged solutions in $F(R)=R+f(R)$ theory shows exact similarity with Einstein gravity in presence of a cosmological constant. Thus in those situations $f(R)$ plays the role of cosmological constant. As shown in \cite{Hendi12} this charged solution is equivalent to Einstein gravity coupled with conformally invariant Maxwell field, such that $f(R)$ term in this situation has the role of electromagnetic field. This acts as a origin of the charge term.

Next we will consider the equilibrium configurations of perfect fluid rotating around this black hole. For notational simplicity we will assume, $M=2$, $Q=1$ and $y=\Lambda/3$. Then the line element for this black hole spacetime reduces to the following form
\begin{equation}\label{FRC05}
ds^{2}=-\left(1-yr^{2}-\frac{2}{r}+\frac{1}{r^{2}}\right)dt^{2}+\frac{dr^{2}}{\left(1-yr^{2}-\frac{2}{r}+\frac{1}{r^{2}}\right)}+r^{2}d\Omega ^{2}
\end{equation}
We must stress that static region exists in this spacetime with sub-critical values of cosmological parameter $y$ as
\begin{equation}\label{FRC06}
y<y_{c}=\frac{1}{16}
\end{equation}
Thus equilibrium configurations are possible in this spacetime provided cosmological parameter $y$ satisfies the constraint given by Eq. (\ref{FRC06}). The equipotential surfaces are being given by
\begin{equation}\label{FRC07}
W(r,\theta)=\ln \left[\frac{r\sin \theta \sqrt{\left(1-yr^{2}-\frac{2}{r}+\frac{1}{r^{2}}\right)}}
{\sqrt{r^{2}\sin ^{2}\theta -\ell ^{2}\left(1-yr^{2}-\frac{2}{r}+\frac{1}{r^{2}}\right)}} \right]
\end{equation}
as well as we have,
\begin{equation}\label{FRC08}
\frac{d\theta}{dr}=\frac{\left(2r-2-2yr^{4}\right)\sin ^{2}\theta-2\ell ^{2}\left(1-yr^{2}-\frac{2}{r}+\frac{1}{r^{2}}\right)^{2}}
{2r\ell ^{2}\left(1-yr^{2}-\frac{2}{r}+\frac{1}{r^{2}}\right)^{2}}\tan \theta
\end{equation}
Note that for $y=0$ it reduces to the RN scenario discussed
earlier. The best idea about nature of $\ell =\textrm{constant}$
surface can be extracted from behaviour of the potential
$W(r,\theta)$ in equatorial plane ($\theta =\pi/2$). There we have
two reality conditions imposed on the potential,
\begin{eqnarray}\label{FRC09}
\left(1-yr^{2}-\frac{2}{r}+\frac{1}{r^{2}}\right)\geq 0\\
\ell ^{2}\leq \ell _{ph}^{2}\equiv \frac{r^{2}}{\left(1-yr^{2}-\frac{2}{r}+\frac{1}{r^{2}}\right)}
\end{eqnarray}
The function $\ell _{ph}$ represents angular momentum of photon's orbit. Further extremizing the potential $W(r,\theta =\pi/2)$ we arrive at the particular expression for angular momentum density
\begin{equation}\label{FRC10}
\ell ^{2}=\ell _{K}^{2}(r,y)\equiv \frac{\left(2r-2-2yr^{4}\right)}{2\left(1-yr^{2}-\frac{2}{r}+\frac{1}{r^{2}}\right)^{2}}
\end{equation}
The specific energy of circular geodesics, corresponding to local extrema of the effective potential can be expressed as
\begin{equation}\label{FRC11}
E_{c}(r,y)=\frac{1-yr^{2}-\frac{2}{r}+\frac{1}{r^{2}}}{\sqrt{1-\frac{3}{r}+\frac{2}{r^{2}}}}
\end{equation}

\begin{figure*}
\begin{center}

(a)
\includegraphics[height=2in, width=2.5in]{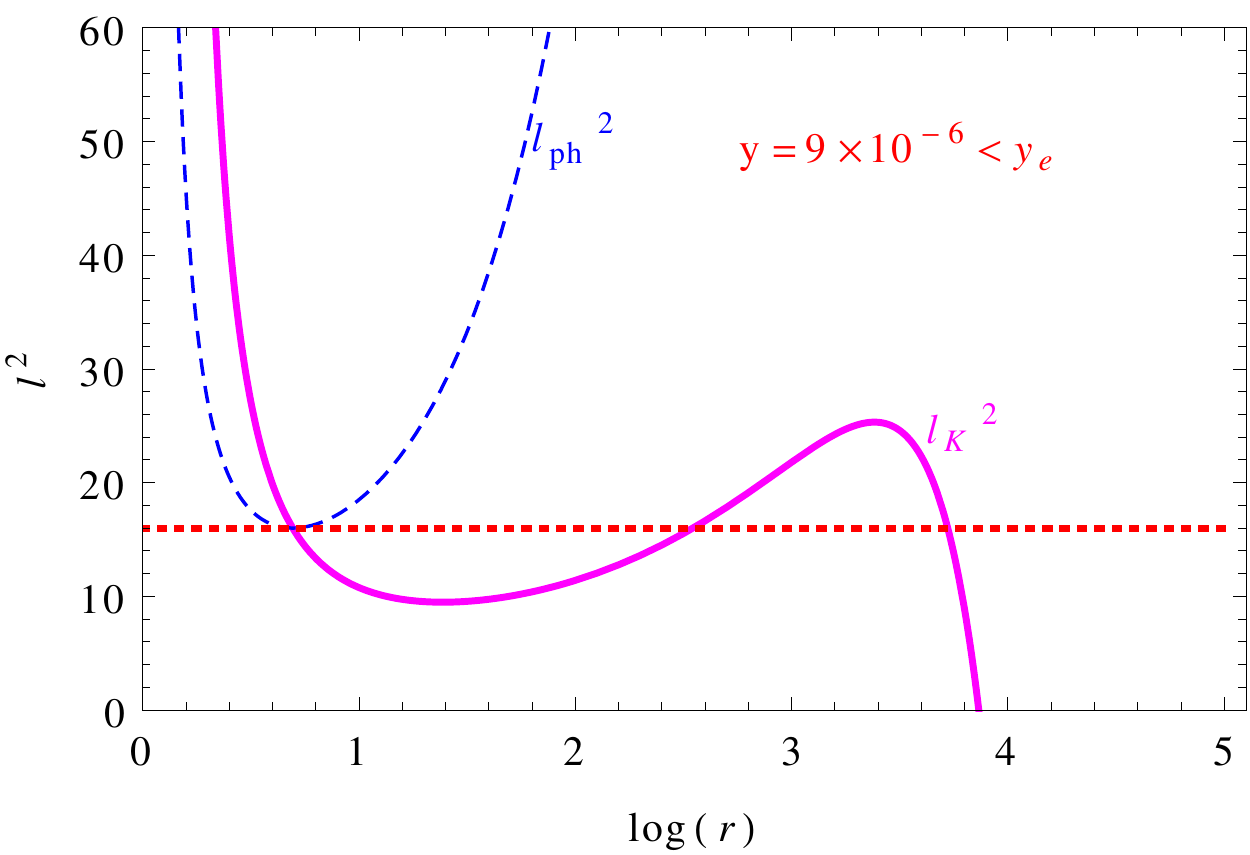}~~~
(b)
\includegraphics[height=2in, width=2.5in]{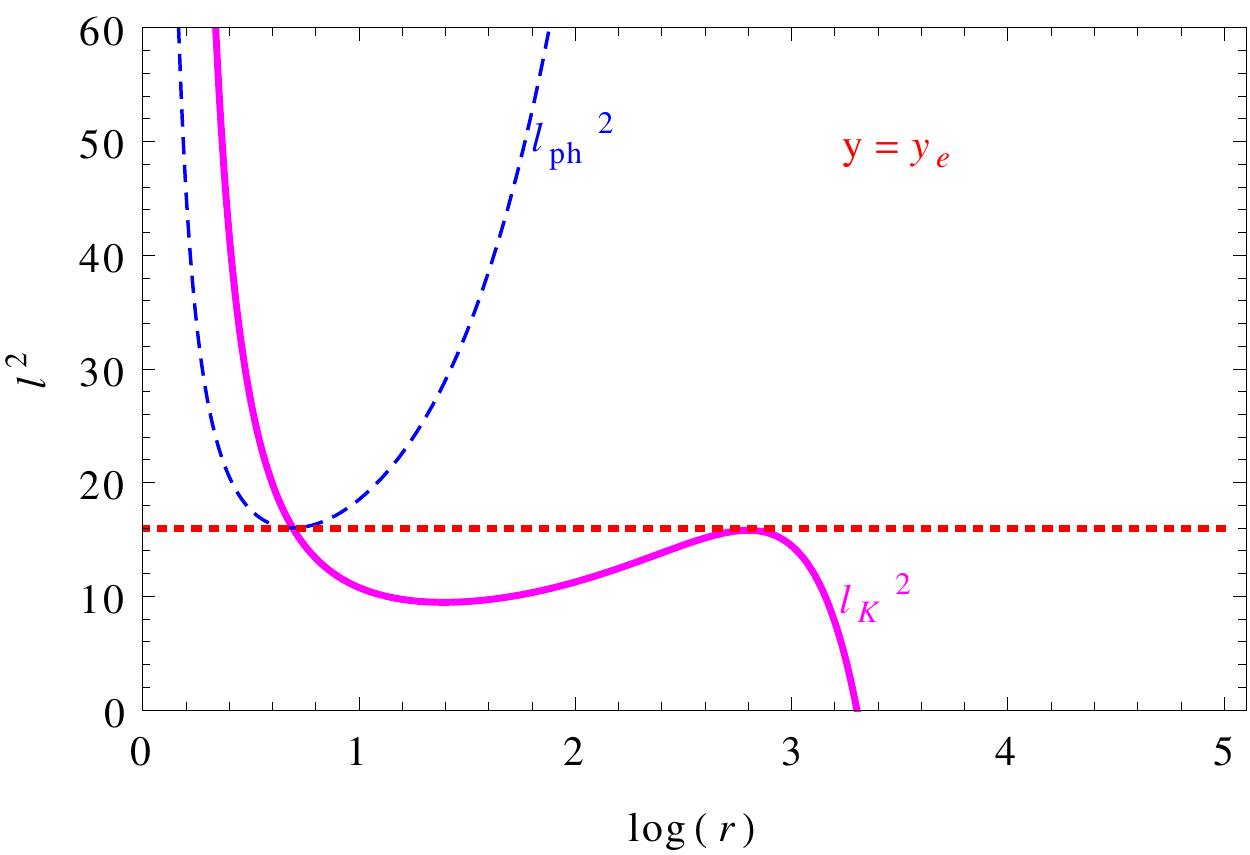}\\
(c)
\includegraphics[height=2in, width=2.5in]{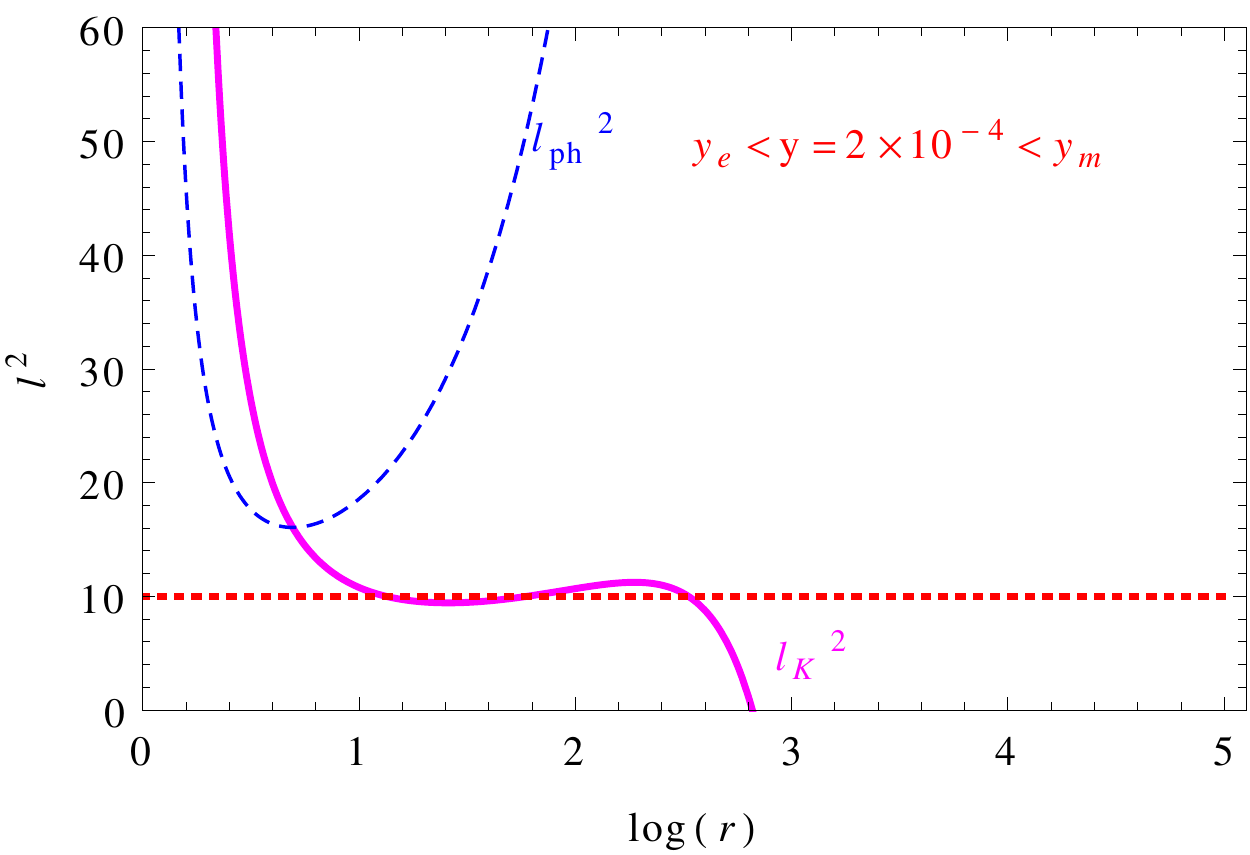}~~~
(d)
\includegraphics[height=2in, width=2.5in]{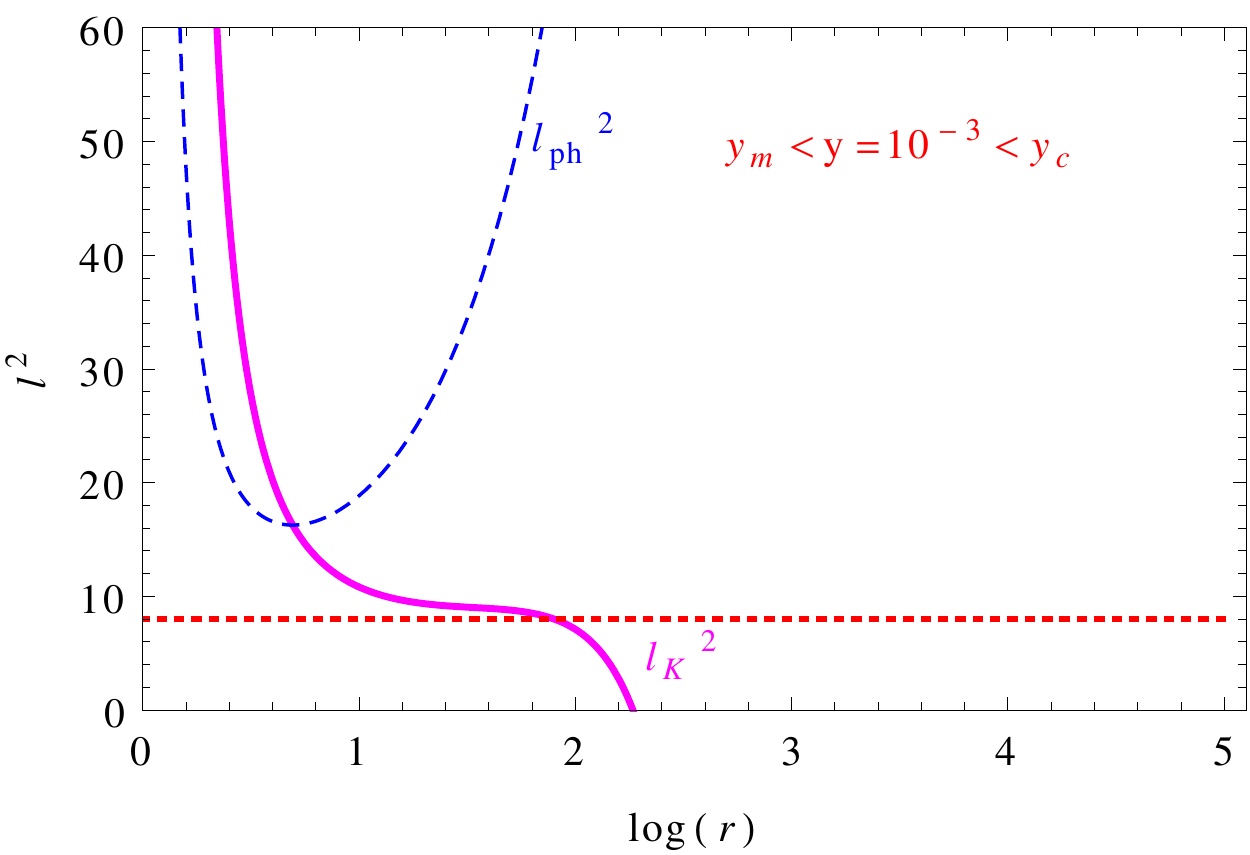}\\

\caption{The figures (a)-(d) show the variation of $l_{ph}^{2}(r,y)$ and $l_{K}^{2}(r,y)$
with radial variable $r$ for four different values of cosmological
parameter $y$ with $M=2$ and $Q=1$
(thus both $\ell ^{2}$ and $r$ are in units of $M$).
They reflect subsequently the following cases:
(a)$0<y<y_{e}$;~(b)$y=y_{e}$;~(c)$y_{e}<y<y_{ms}$ and
(d)$y_{ms}<y<y_{c}$. The descending parts of $\ell _{K}^{2}$ determine the
cusps and the growing part determines the central rings. \label{fig3} }

\end{center}
\end{figure*}
Now we turn back to the horizons in this spacetime. For $y>0$,
the photon angular momentum $\ell _{ph}^{2}(r,y)$ diverges at the black hole
horizon $r_{h}$ and cosmological horizon $r_{c}$ determined by the
equality in Eq. (\ref{FRC09}). This leads to the following expression for
horizons,
\begin{eqnarray}\label{FRC12}
r_{h}=\frac{1}{2}\left(\frac{1}{\sqrt{y}}-\frac{\sqrt{1-4\sqrt{y}}}{\sqrt{y}}\right)\\
r_{c}=\frac{1}{2}\left(\frac{1}{\sqrt{y}}+\frac{\sqrt{1-4\sqrt{y}}}{\sqrt{y}}\right)
\end{eqnarray}
Also the photon circular orbit radius can be obtained from the extrema of $\ell _{ph}^{2}$. This has the numerical value: $r_{ph}=3.56155$. For this circular orbit the photon angular momentum density could be given by,
\begin{equation}\label{FRC13}
\ell ^{2} _{ph(c)}\equiv \ell ^{2}_{ph}(r_{ph},y)= \frac{12.6846}{0.359611-12.9846y}
\end{equation}
Also the zero point of $\ell _{K}^{2}$ leads to a radius called
static radius and in this black hole spacetime is expressed
parametrically as,
\begin{equation}\label{FRC14}
y(r_{s})=\frac{r_{s}-1}{r_{s}^{4}}
\end{equation}
Note that $\ell _{K}^{2}$ is not well defined in the region $r>r_{s}$,
being negative there. At this static radius black hole attraction is
compensated for cosmological repulsion. Then local extrema of
$\ell_{K}^{2}$ corresponds to the following expression for cosmological
parameter as,
\begin{equation}\label{FRC15}
y_{ms}(r)=\frac{(r-1)^{4}(r-4)}{r^{4}(12-15r+4r^{2})}
\end{equation}
This determines the marginally stable circular geodesics.
 The local maxima of this function $y_{ms}$ gives the critical
value for the cosmological parameter $y$ admitting stable circular orbits,
\begin{equation}\label{FRC16}
y_{ms}=0.000692
\end{equation}
For $y<y_{ms}$, there exists an inner or outer marginally stable
circular geodesic at $r_{ms(i)}$ or $r_{ms(o)}$.
There exists another special value of $y$, which corresponds
to the situation where minimum value of $\ell _{ph}^{2}$
equals the maximum of $\ell _{K}^{2}$. This value is denoted
by $y_{e}$ and has the following numerical value,
\begin{equation}\label{FRC17}
y_{e}\sim 4.8\times 10^{-5}
\end{equation}

\begin{figure*}
\begin{center}

(a1)
\includegraphics[height=2in, width=2.5in]{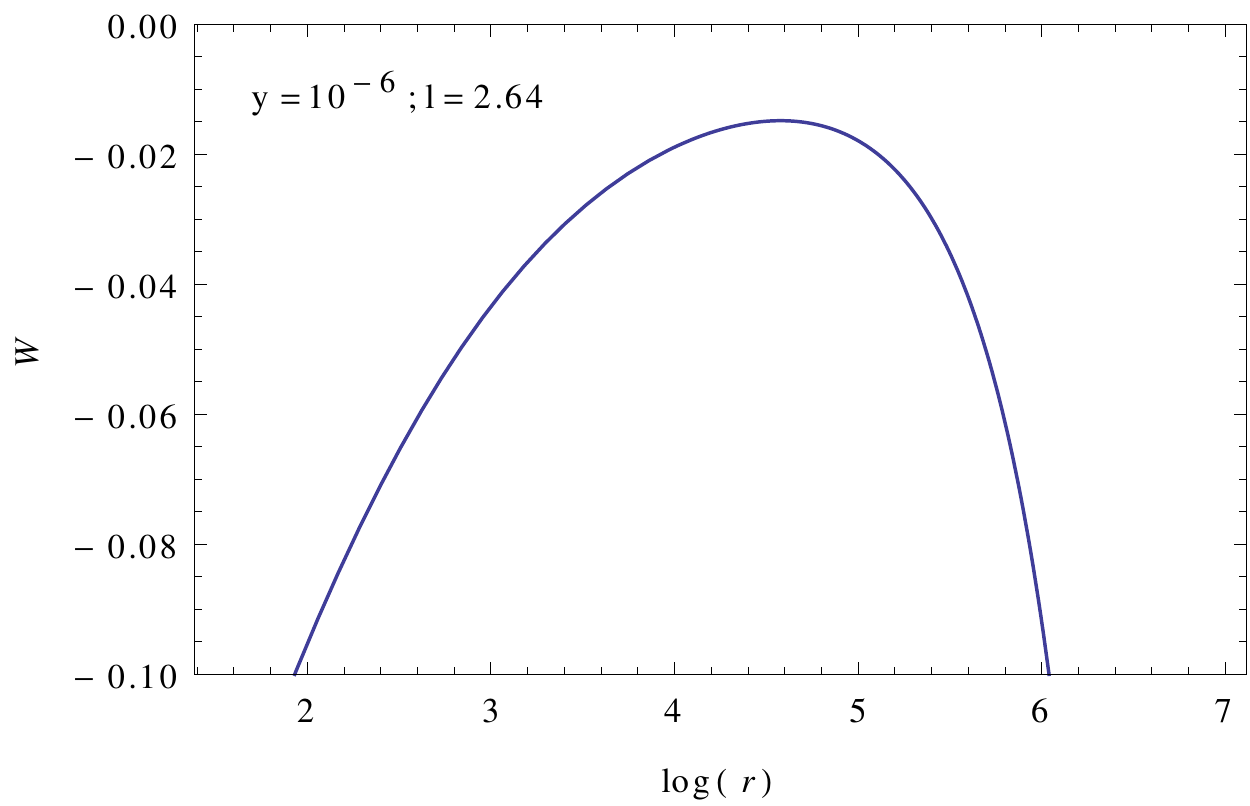}~~~
(a2)
\includegraphics[height=2in, width=2.5in]{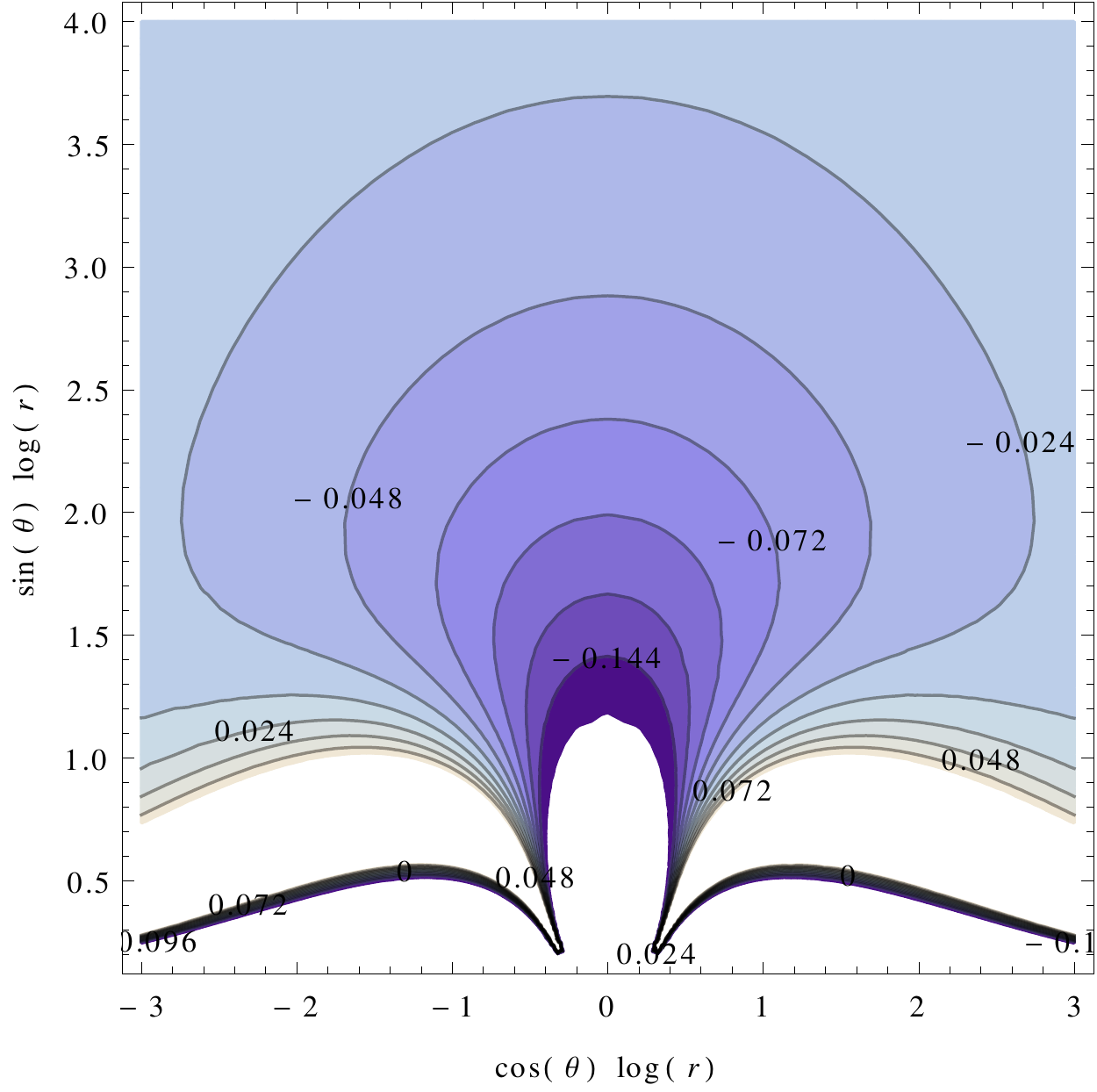}\\
(b1)
\includegraphics[height=2in, width=2.5in]{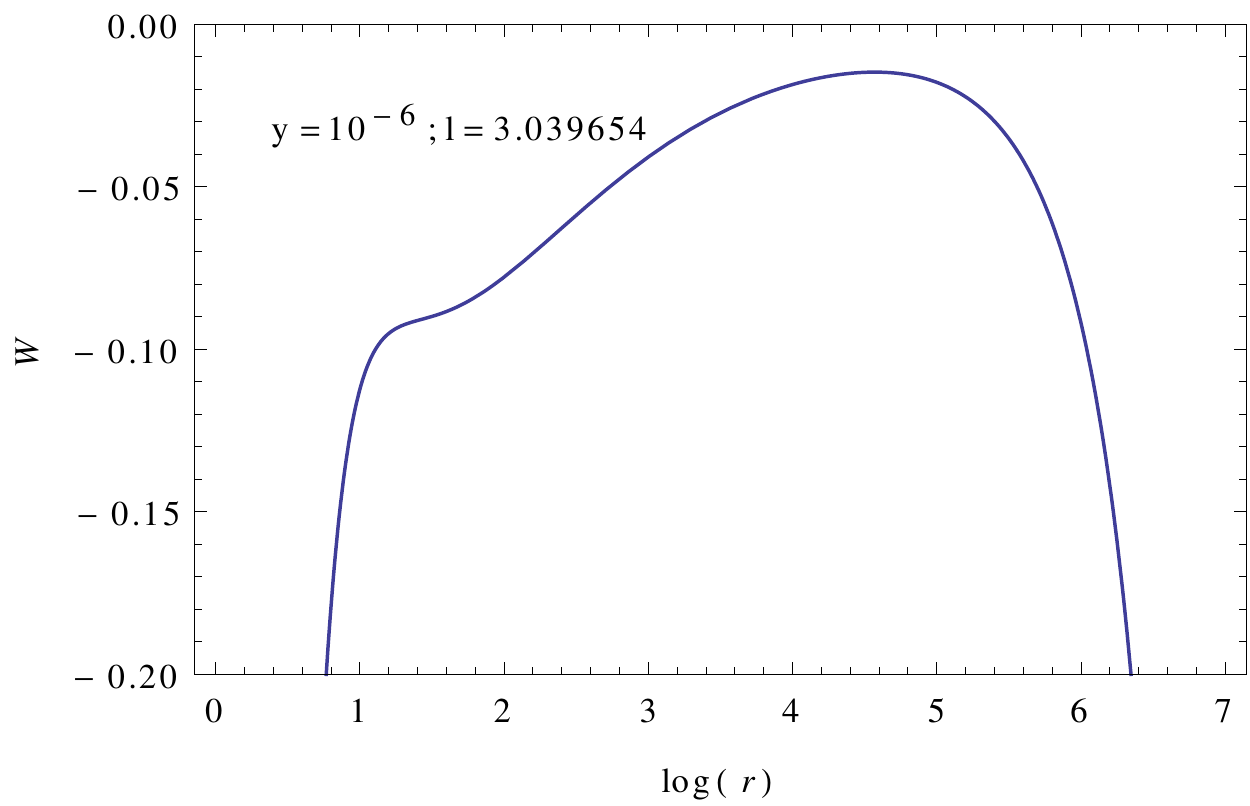}~~~
(b2)
\includegraphics[height=2in, width=2.5in]{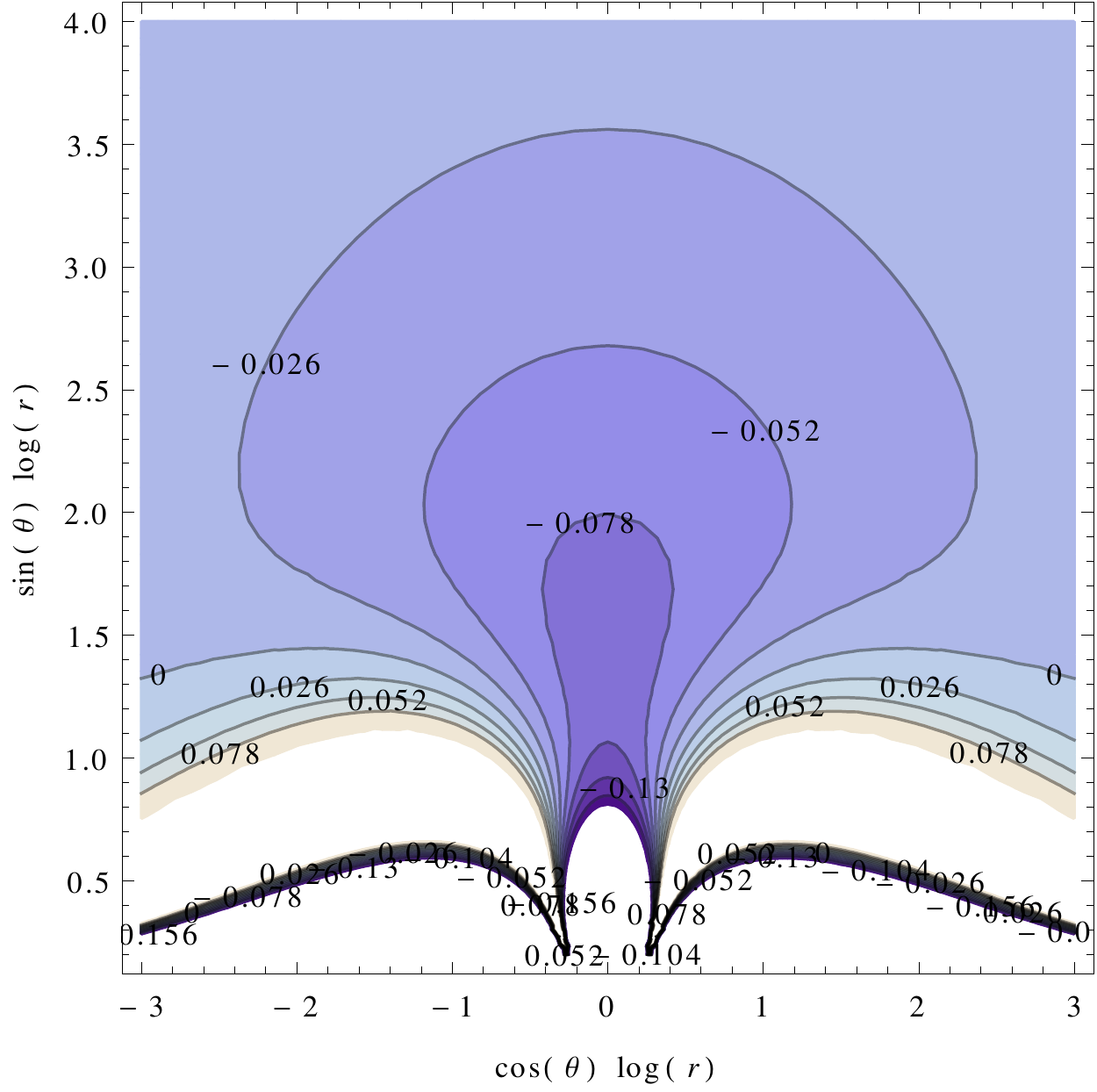}\\
(c1)
\includegraphics[height=2in, width=2.5in]{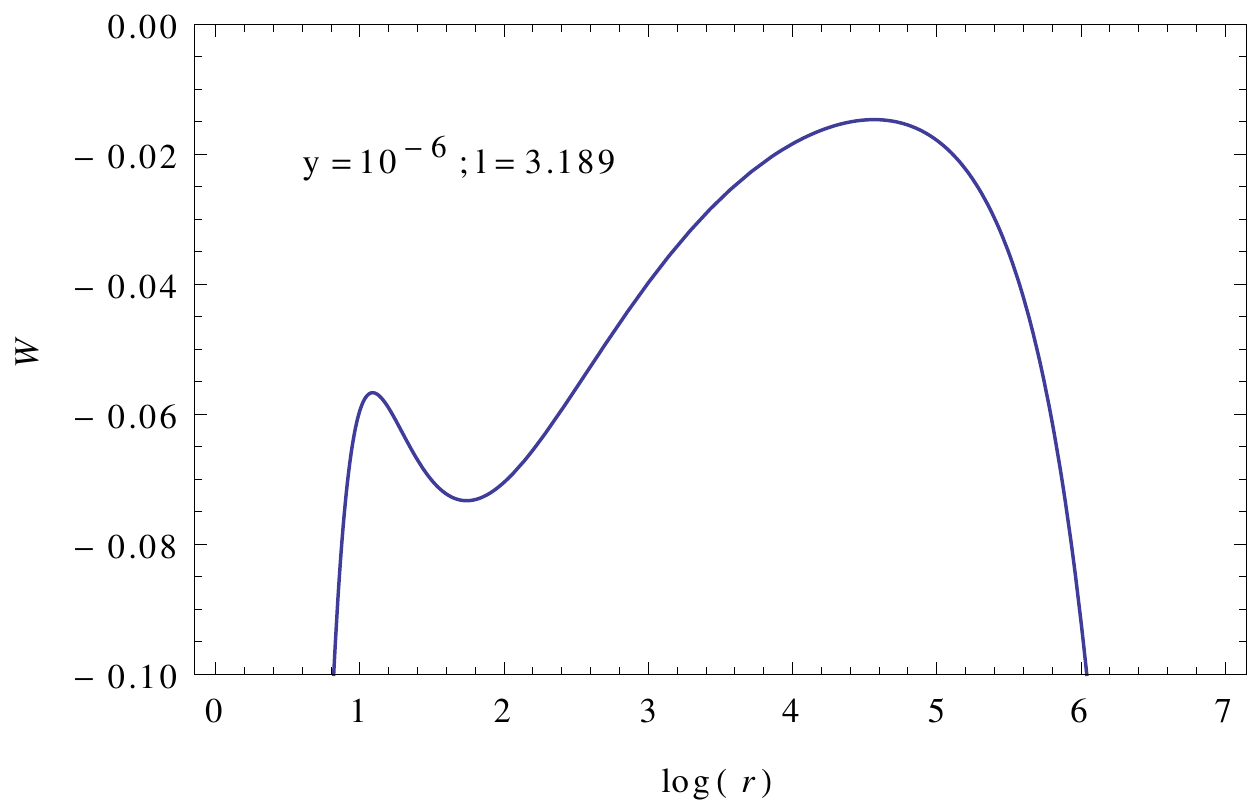}~~~
(c2)
\includegraphics[height=2in, width=2.5in]{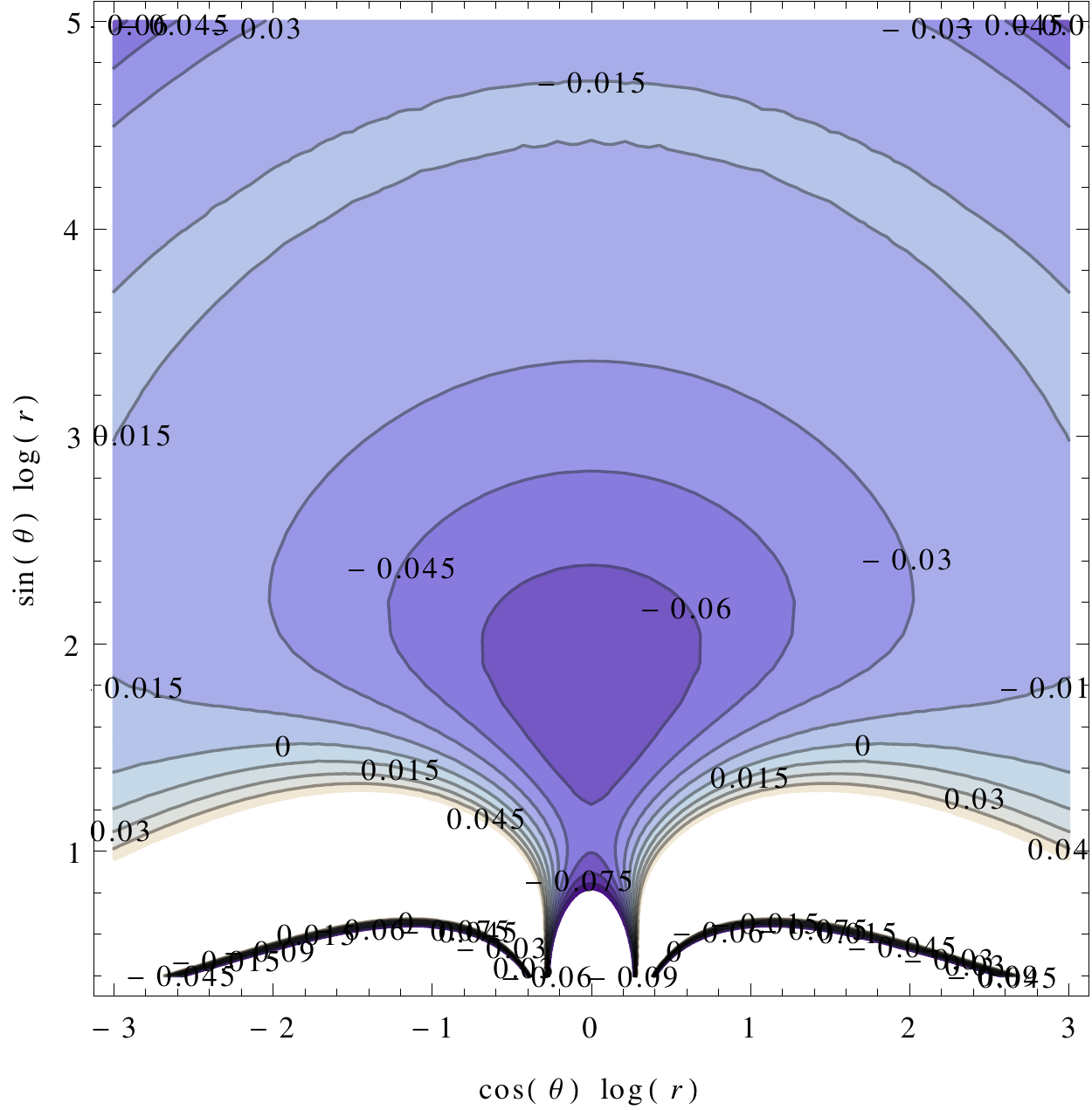}\\
(d1)
\includegraphics[height=2in, width=2.5in]{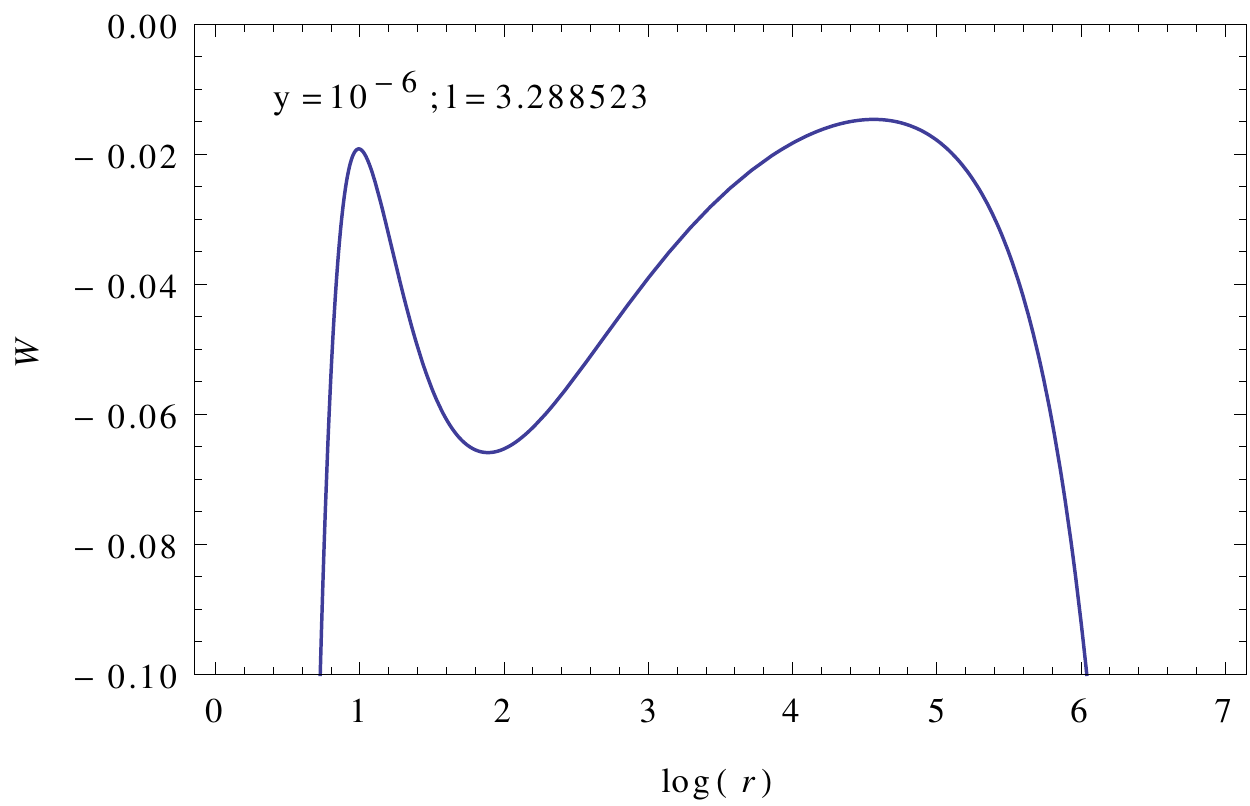}~~~
(d2)
\includegraphics[height=2in, width=2.5in]{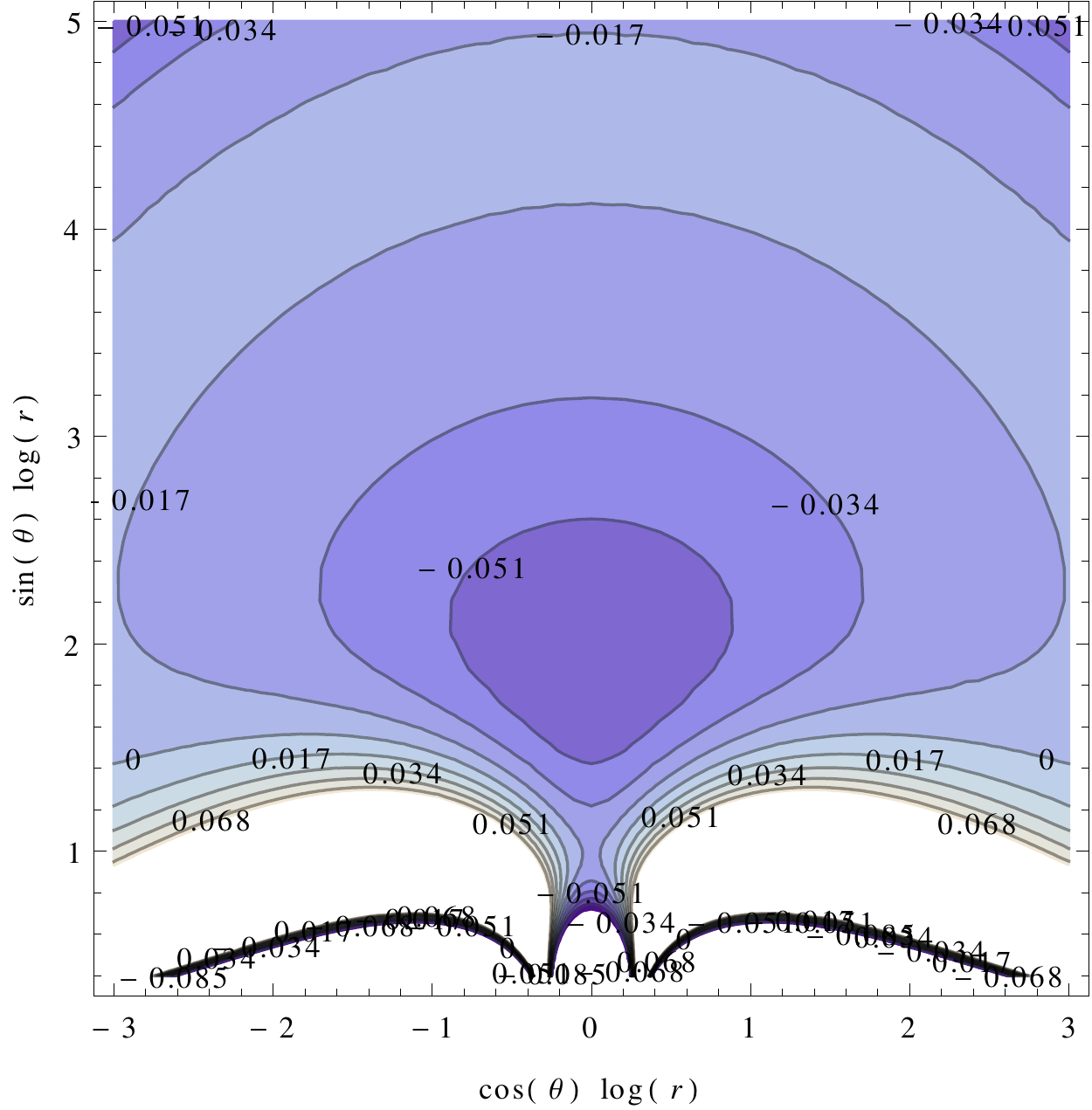}\\

\end{center}
\end{figure*}
\begin{figure*}
\begin{center}

(e1)
\includegraphics[height=2in, width=2.5in]{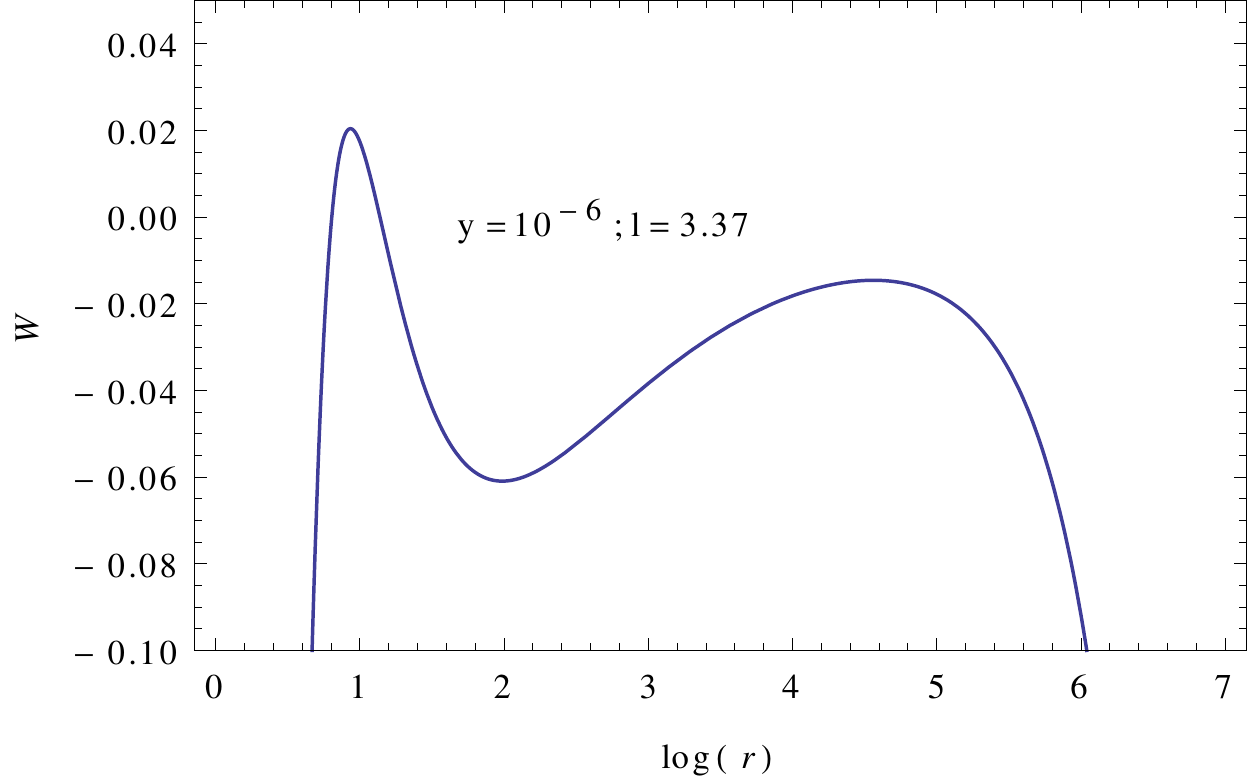}~~~
(e2)
\includegraphics[height=2in, width=2.5in]{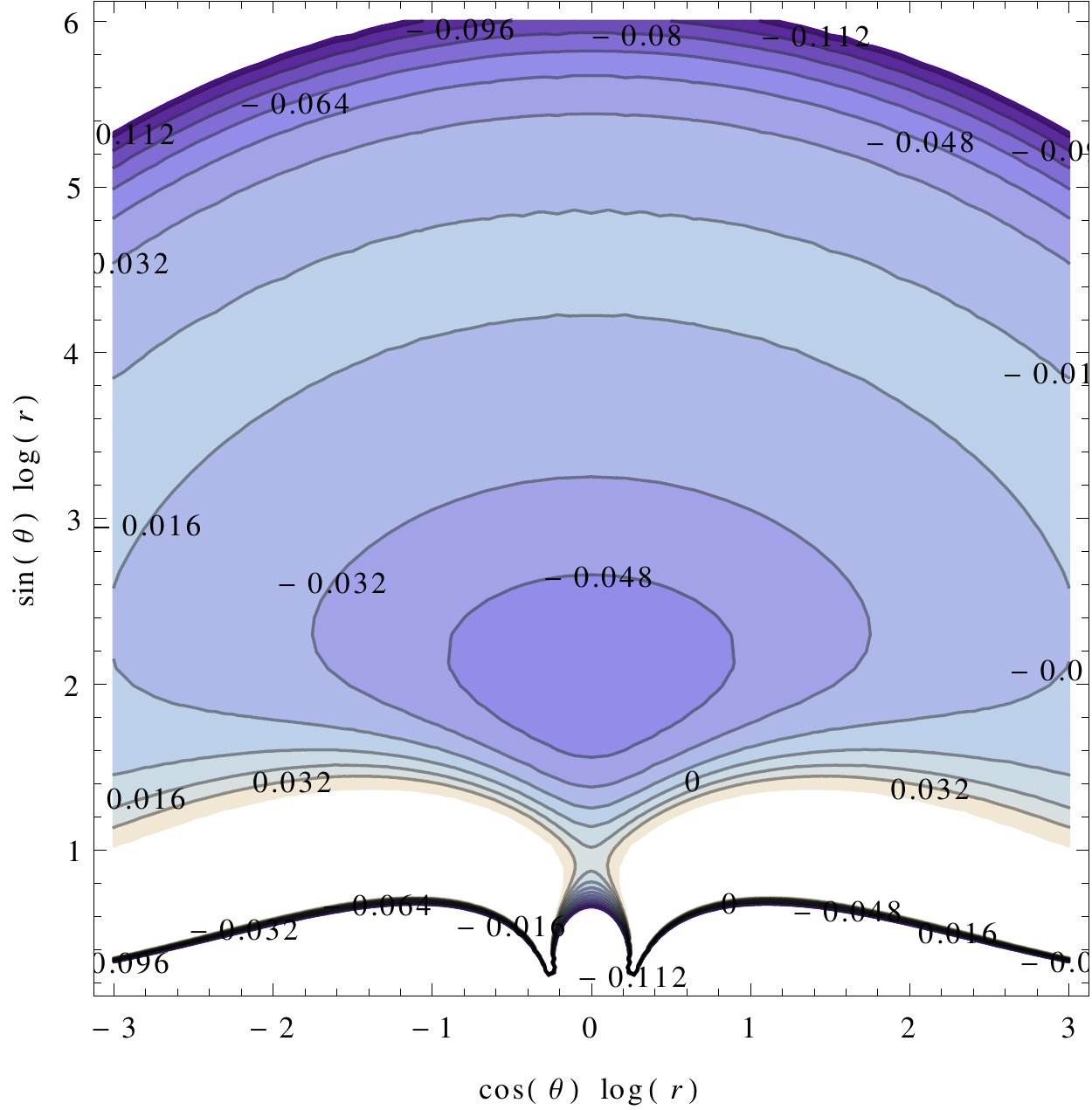}\\
(f1)
\includegraphics[height=2in, width=2.5in]{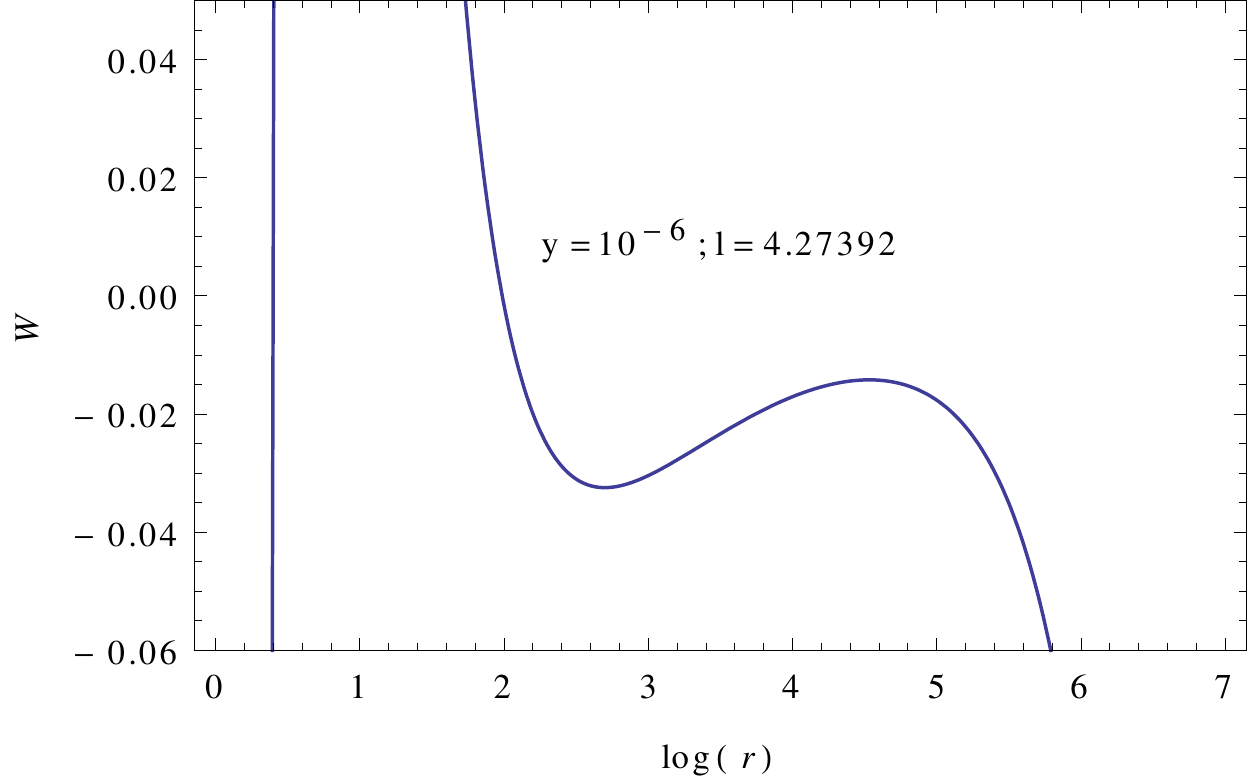}~~~
(f2)
\includegraphics[height=2in, width=2.5in]{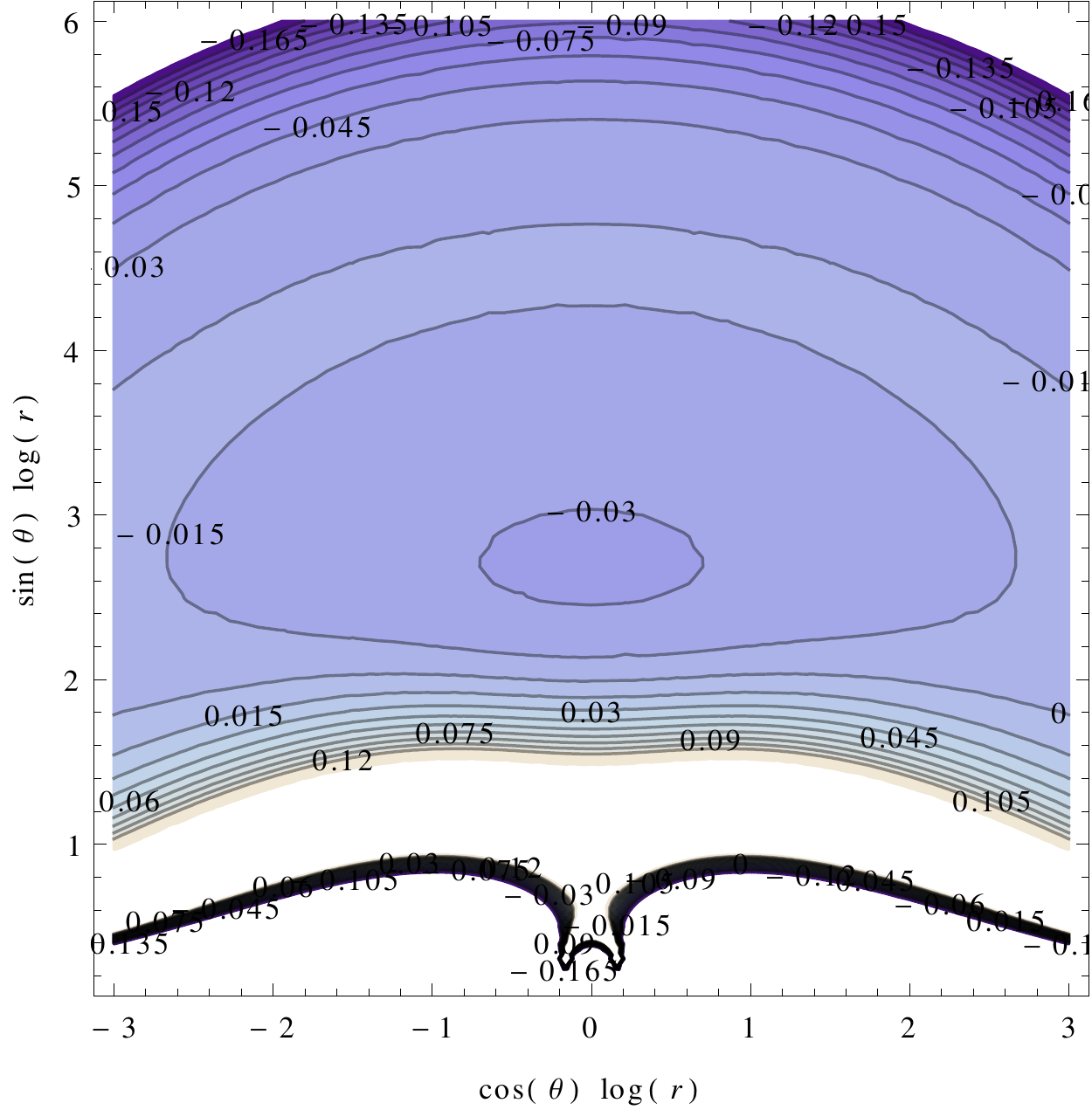}\\
(g1)
\includegraphics[height=2in, width=2.5in]{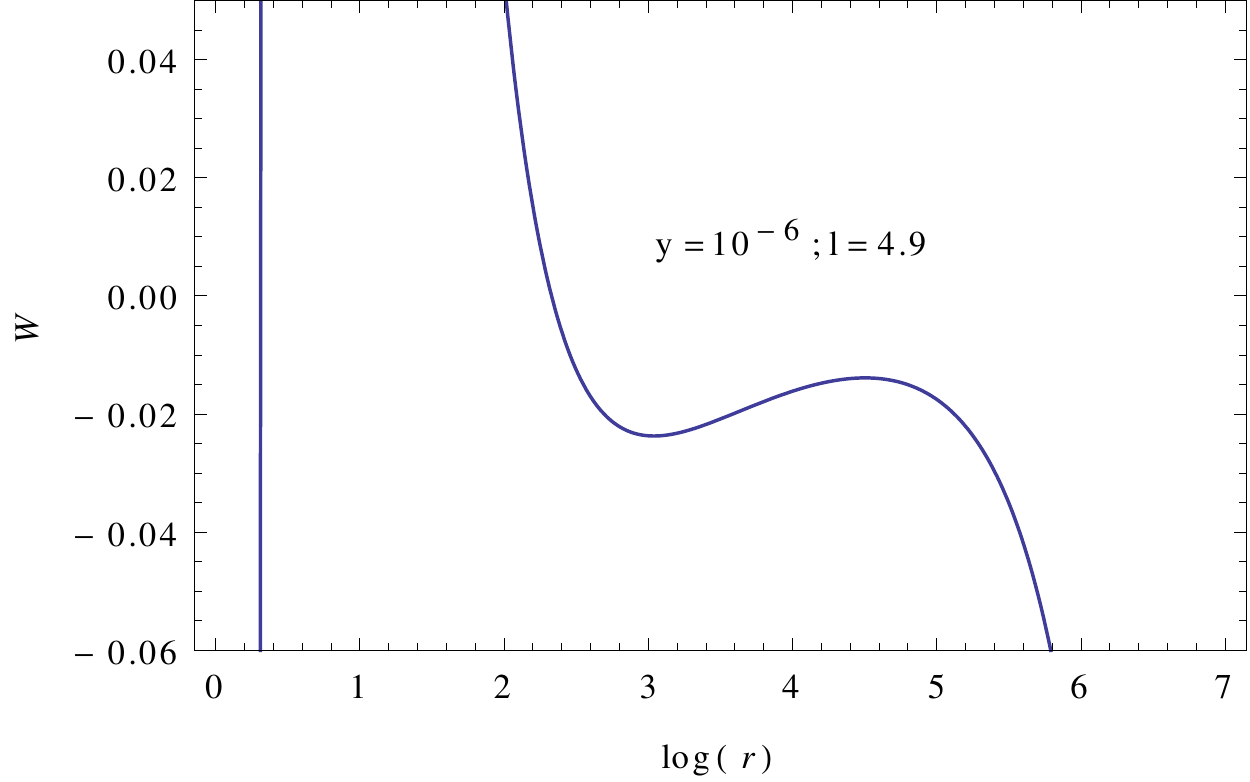}~~~
(g2)
\includegraphics[height=2in, width=2.5in]{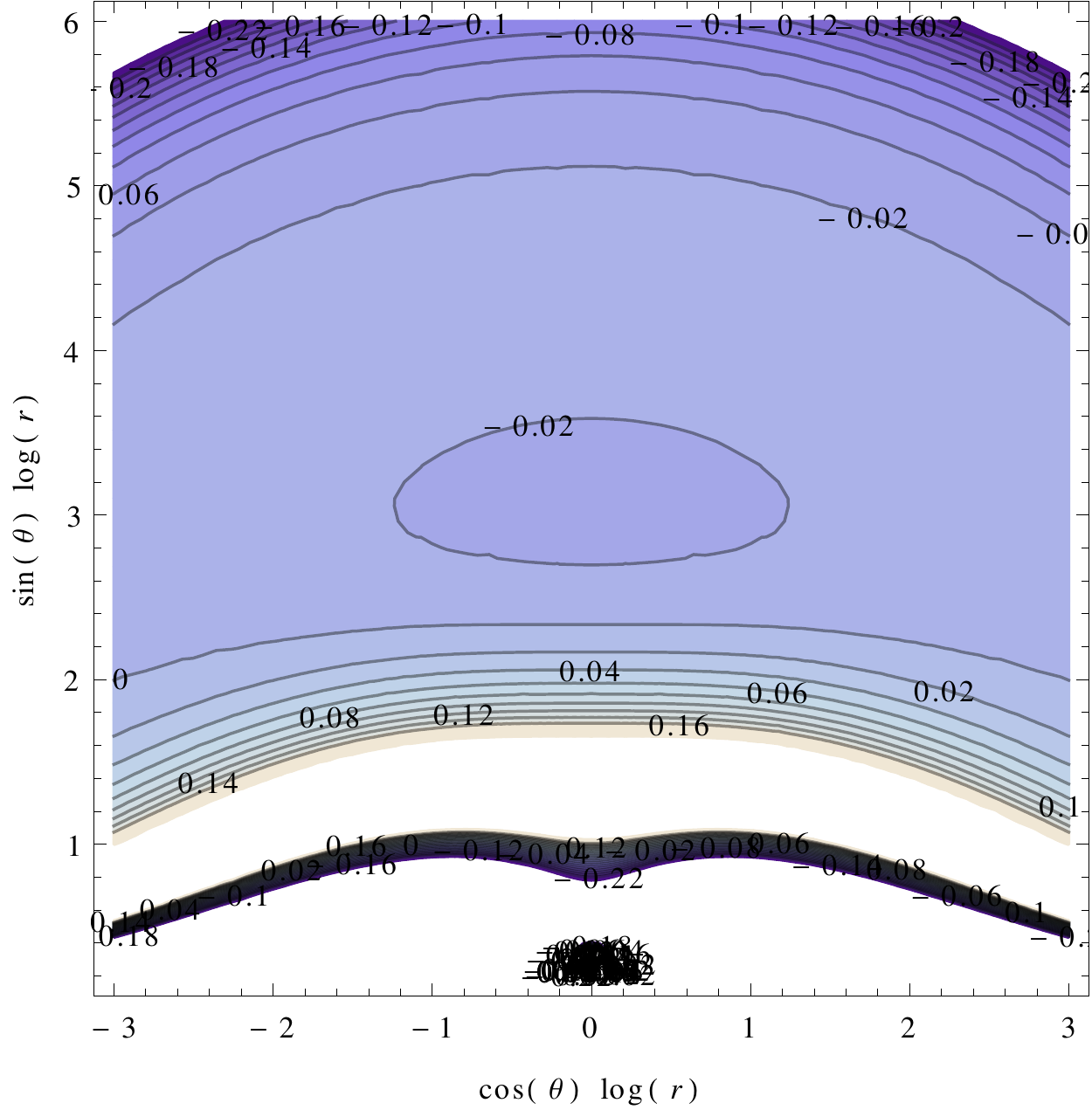}\\
(h1)
\includegraphics[height=2in, width=2.5in]{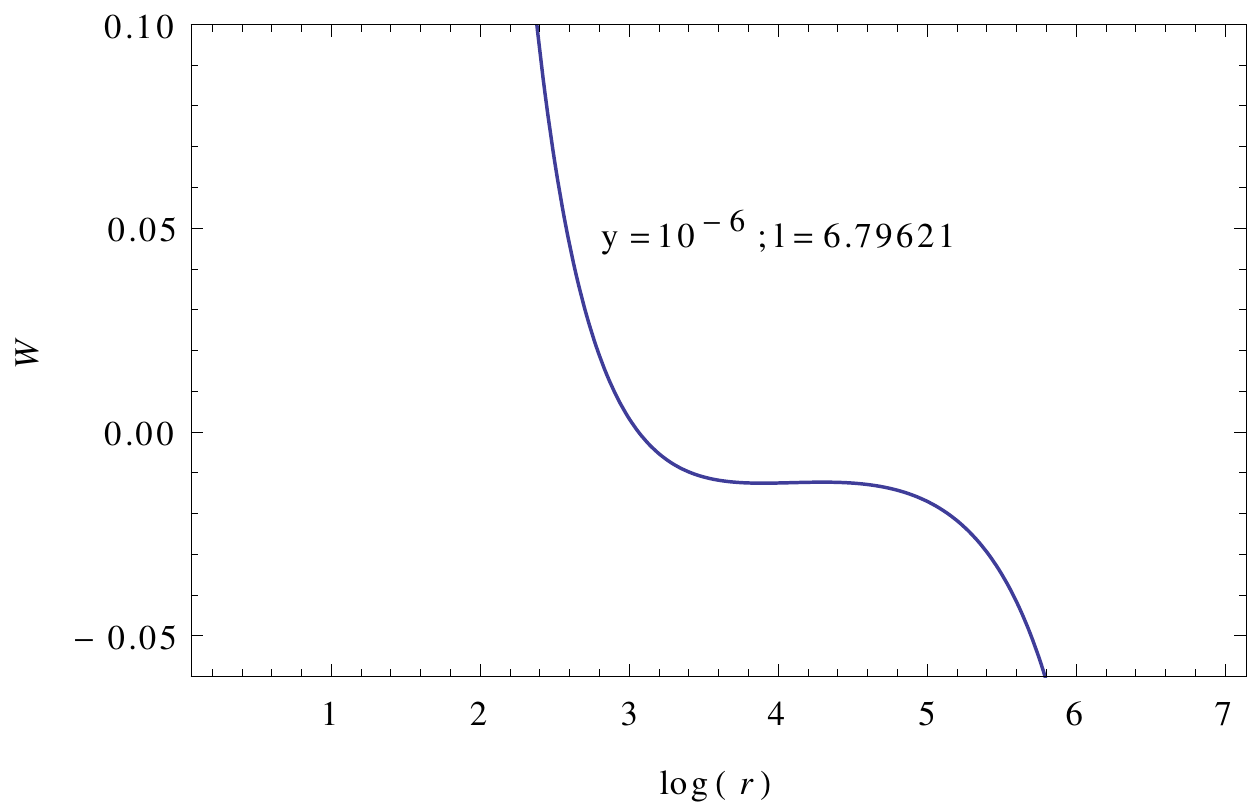}~~~
(h2)
\includegraphics[height=2in, width=2.5in]{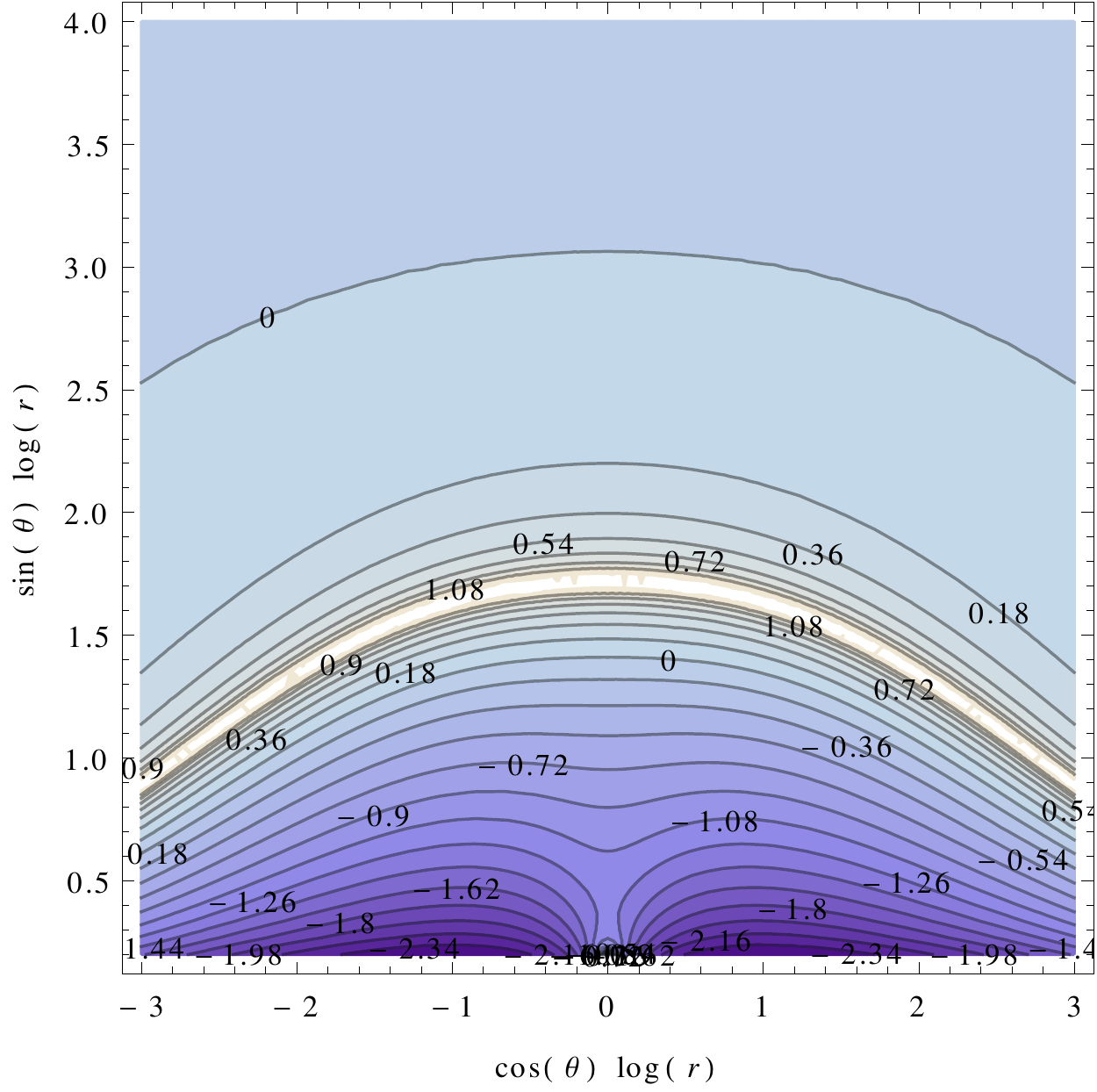}\\

\end{center}
\end{figure*}

\begin{figure*}
\begin{center}

(i1)
\includegraphics[height=2in, width=2.5in]{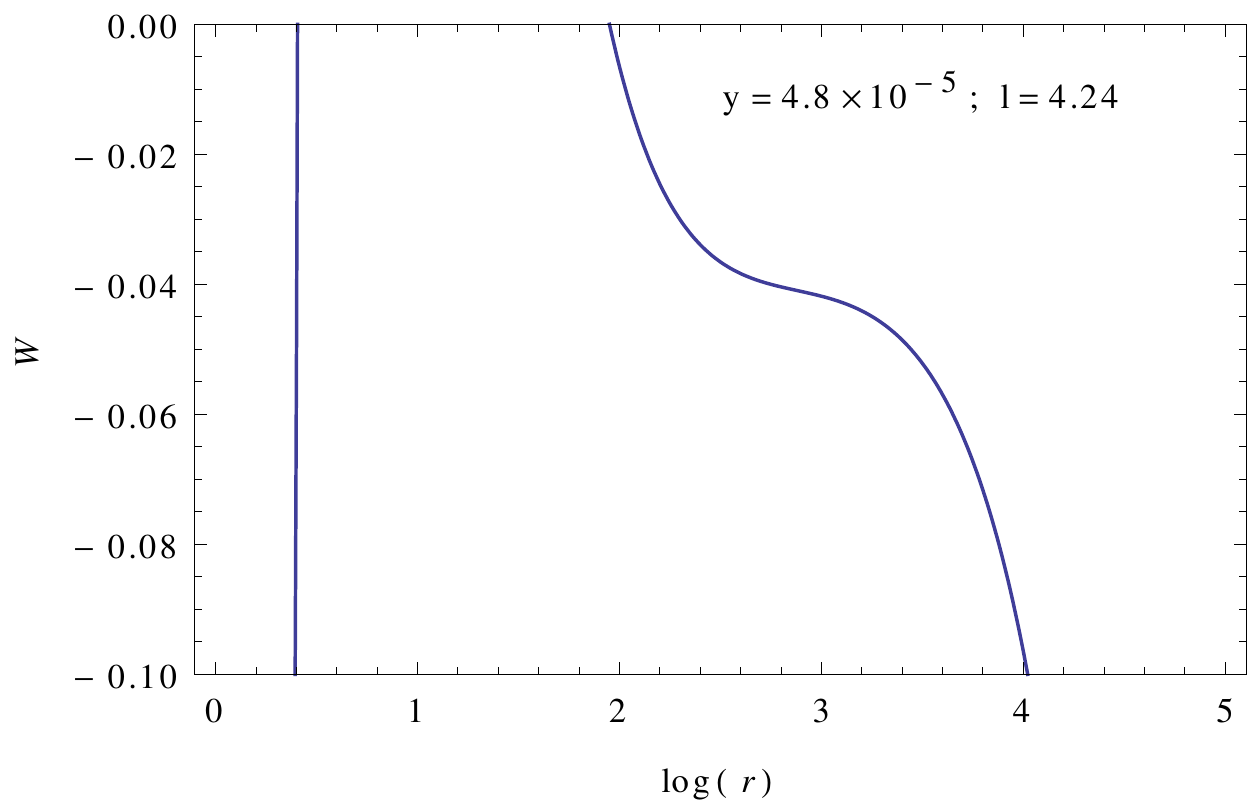}~~~
(i2)
\includegraphics[height=2in, width=2.5in]{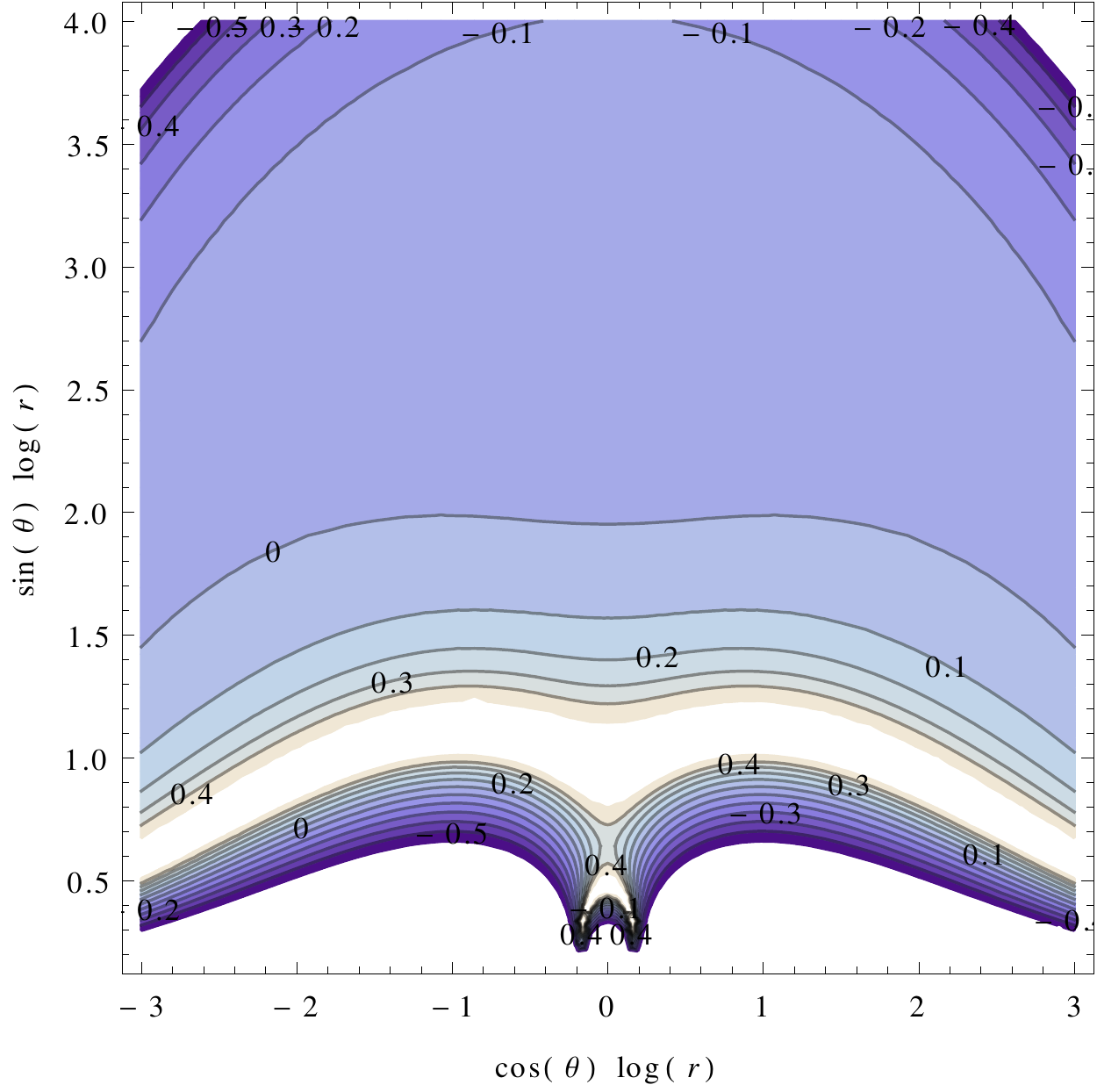}\\
(j1)
\includegraphics[height=2in, width=2.5in]{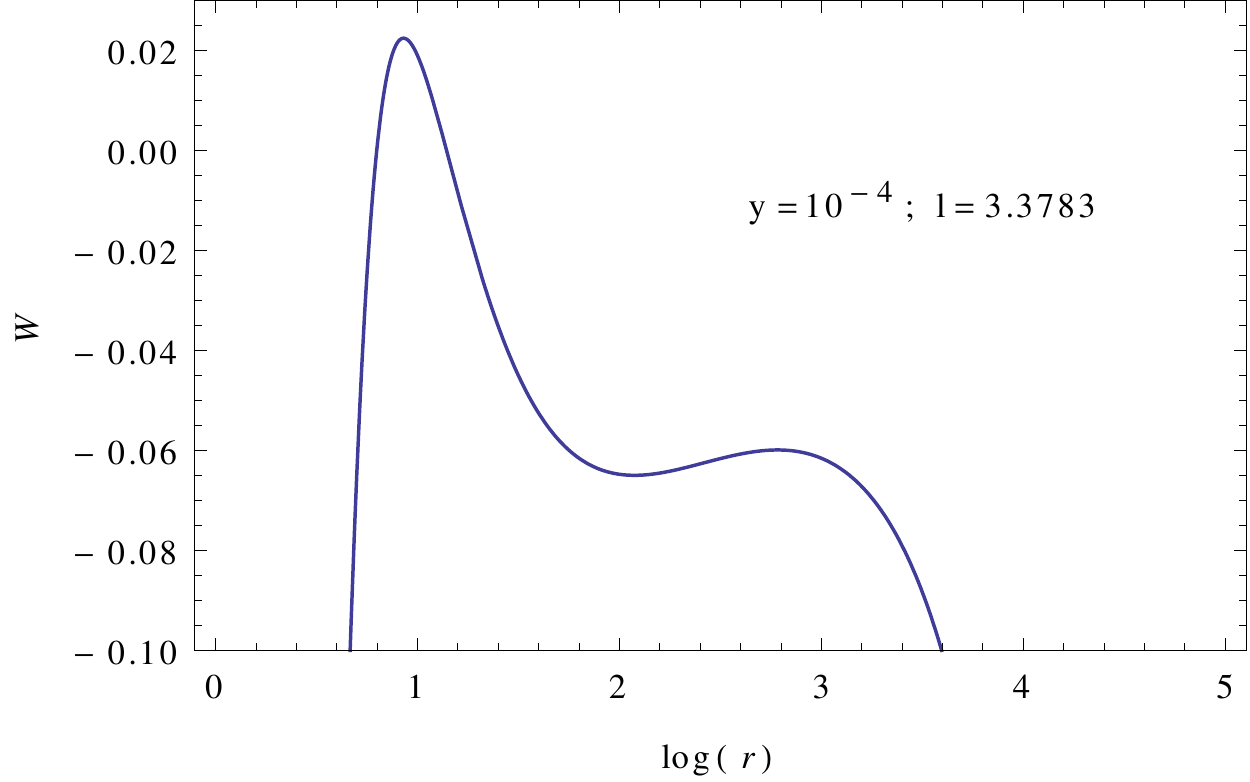}~~~
(j2)
\includegraphics[height=2in, width=2.5in]{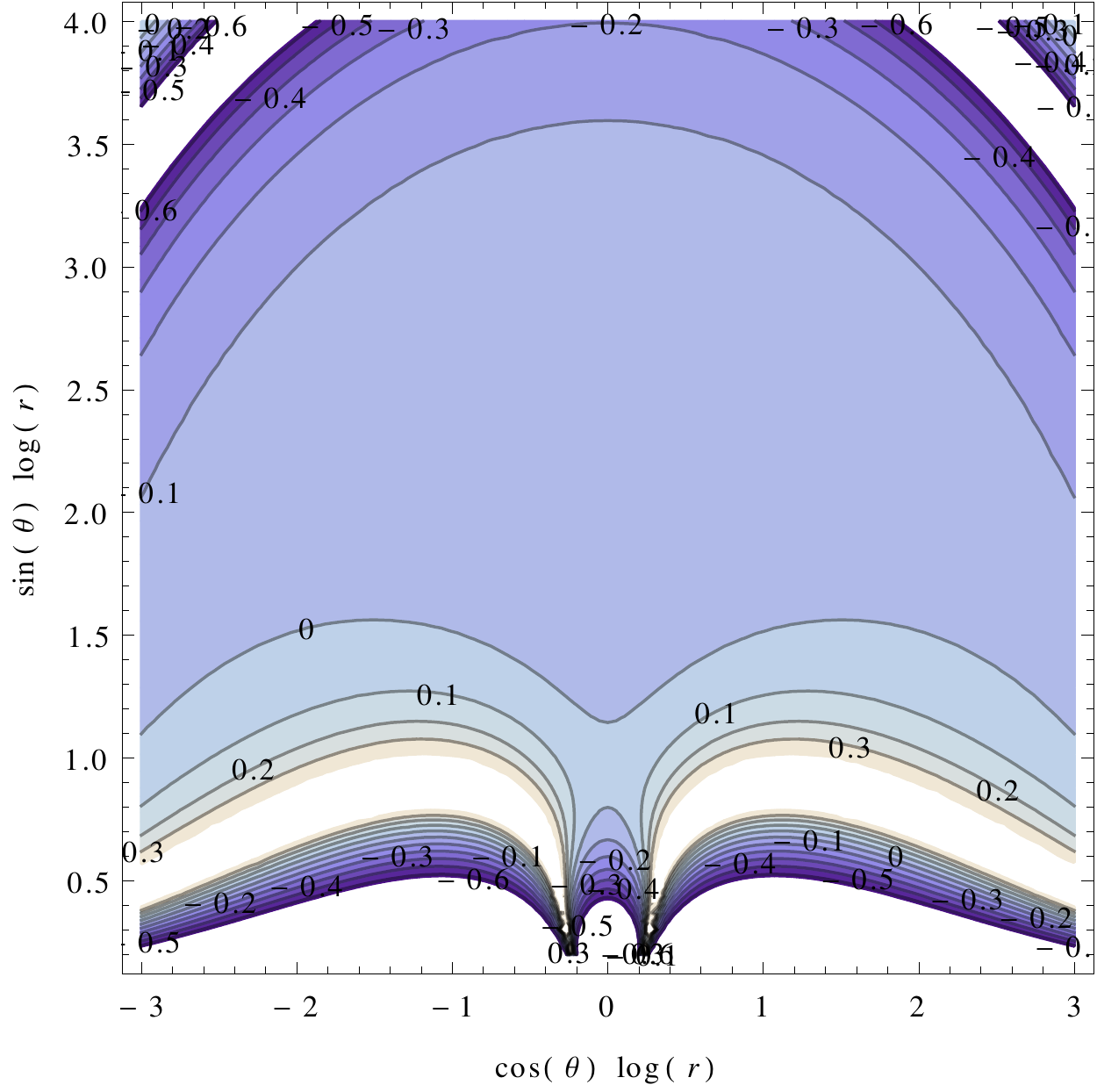}\\
(k1)
\includegraphics[height=2in, width=2.5in]{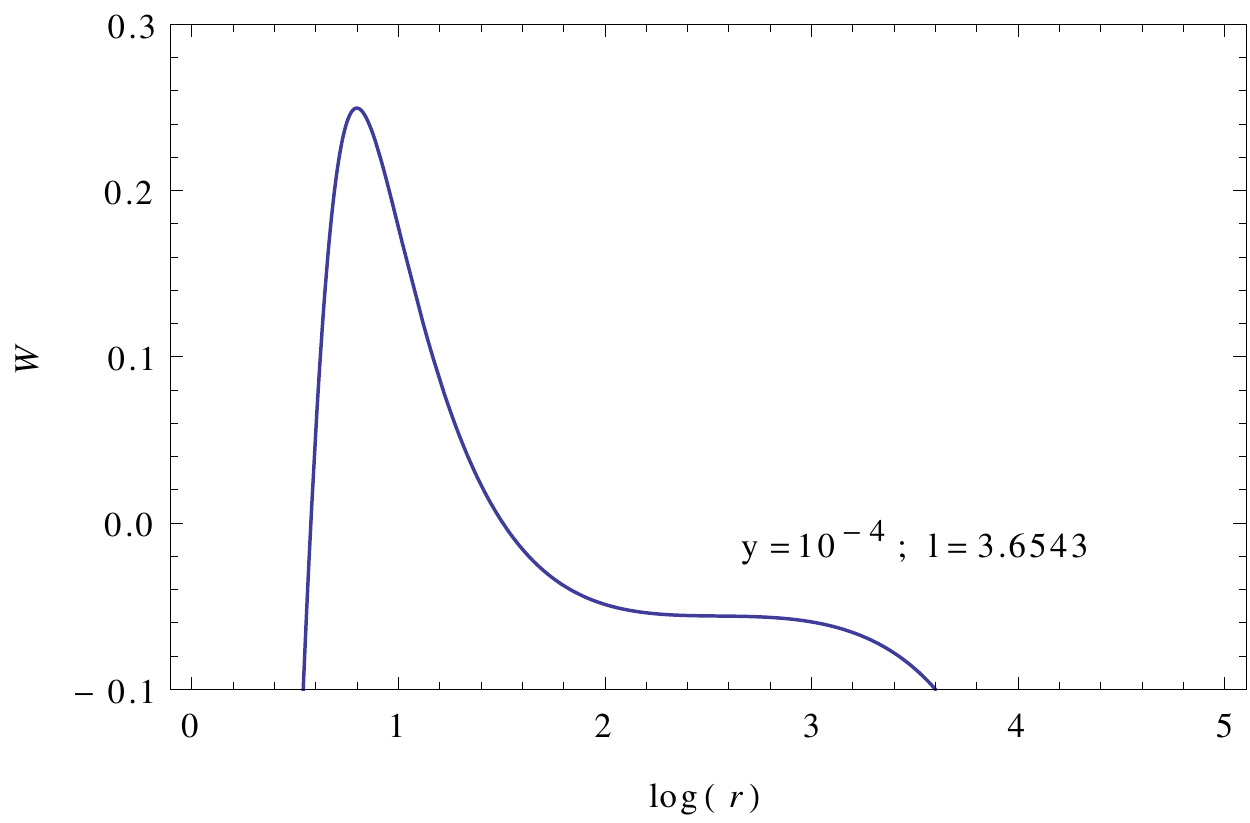}~~~
(k2)
\includegraphics[height=2in, width=2.5in]{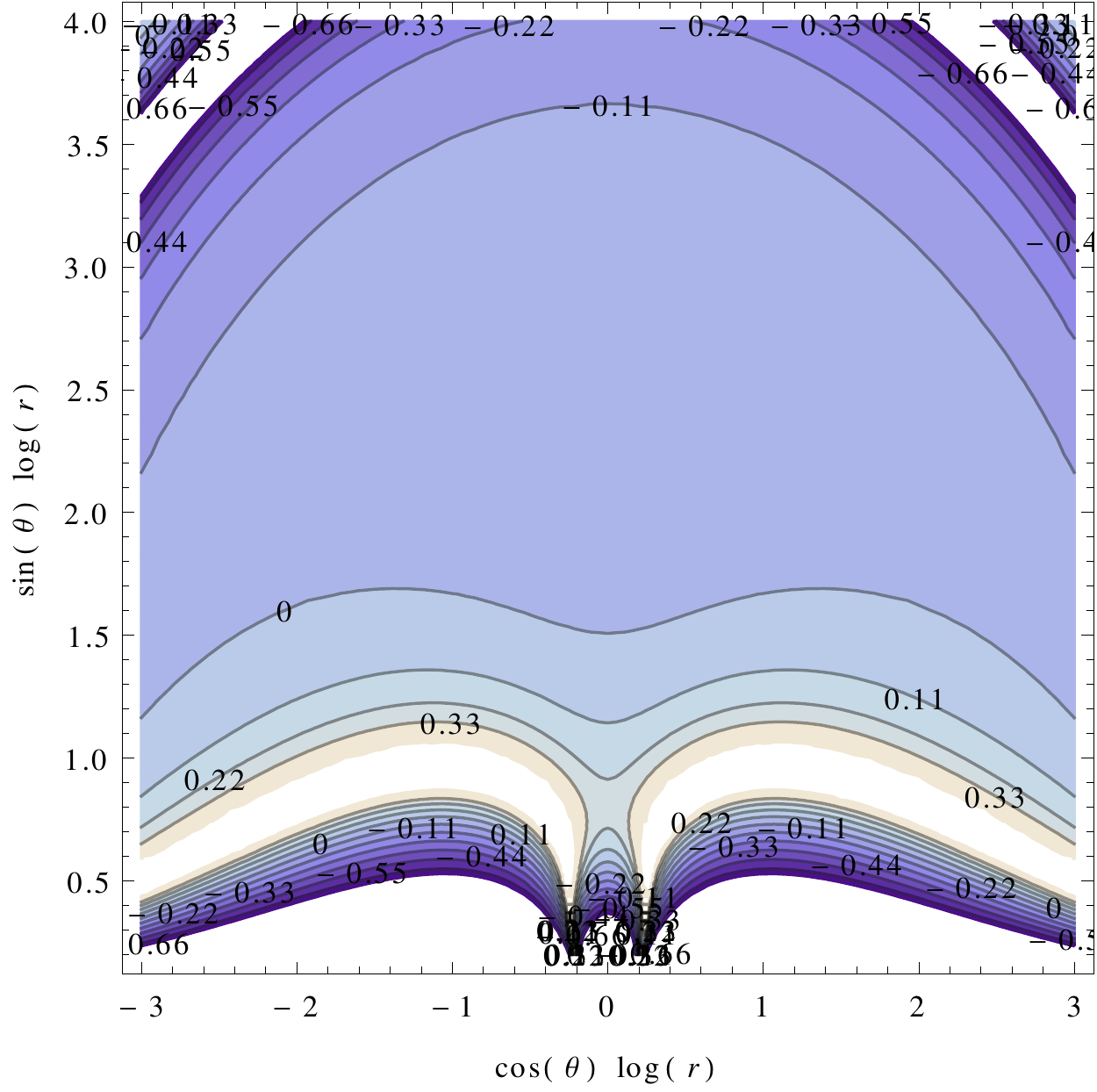}\\

\caption{The figures (a)-(k) are showing the
variation of potential $W(r,\theta=\pi/2,y)$ with $log r$ (the `1' part) and the contour plot
of $W$ with $\log r~\sin \theta$ and $\log r~\cos \theta$ (the `2' part).
These actually determines the equipotential surfaces i.e. meridional sections of
marginally stable configurations with $\ell =\textrm{constant}$ for fluid motion
around black hole in this $F(R)$ theory.
All (a)-(k) figure shows all possible variations for different
values of cosmological parameter $y$ and angular momentum $\ell$.
The cusps are originating from local maxima of the potential
while the central rings are from local minima. All these
situations are more elaborately discussed in the text. \label{fig4} }

\end{center}
\end{figure*}

We can now separate out four different classes depending on the
behavior of the functions $\ell _{ph}^{2}$ and $\ell _{K}^{2}$.
These four classes are defined according to the cosmological parameter in the following way,
\begin{enumerate}
\item $0<y<y_{e}$
\item $y=y_{e}$
\item $y_{e}<y<y_{ms}$
\item $y_{ms}<y<y_{c}$
\end{enumerate}
For all these classes the variation of the two angular momentum
quantities are illustrated in Fig. \ref{fig3}.
In all these figures the descending parts of the curve
$\ell _{K}^{2}(r,y)$ connects to unstable circular geodesics,
along with the growing part (if exists) determining stable
circular geodesics. The extrema of $\ell _{K}^{2}(r,y)$ has an important
significance: the minima determines $\ell _{ms(i)}$ at $r_{ms(i)}$,
the inner marginally stable circular orbit
and the maxima determines $\ell _{ms(o)}$ at $r_{ms(o)}$,
which corresponds to outer marginally stable circular orbit.

The equipotential surfaces are also astrophysically quiet
important. Their properties can be established by studying the
properties of the potential $W(r,\theta)$ in the equatorial plane.
Note that the potential $W(r,\theta =\pi /2,y)$ has closely
related properties with effective potential of geodesic motion. It
is worthwhile to mention that in the limit $r\rightarrow
r_{h}$ or $r\rightarrow r_{c}$ $W(r,\theta =-\pi/2,y)\rightarrow
-\infty$ hence the topological properties of the equipotential
surfaces are directly inferred from the behavior of the potential
$W(r,\theta =\pi/2,y)$. As pointed out earlier, local extrema of
the potential is determined by the condition,
\begin{equation}
\ell ^{2}=\ell ^{2}_{K}(r,y)
\end{equation}
Hence the decreasing part of $\ell _{K}^{2}$ determines the maxima
of the potential $W(r,\theta =\pi/2,y)$. These correspond to cusps
such that at these radii matter moves along unstable geodesic orbit.
While the rising part in $\ell _{K}^{2}$ determines the minima of the
potential. These correspond to central rings of the equilibrium
configurations along which matter moves in stable geodesics.

Next we provide a complete study of the behavior of equipotential
surfaces along with the potential $W(r,\theta =\pi/2,y)$.
We begin with the situations of astrophysical importance.
\begin{enumerate}

\item $~0<y<y_{e}$. This is the situation illustrated in Fig.\ref{fig3}a.
Here we discuss eight different
configurations according to the values of $\ell =\textrm{constant}$ satisfying $\ell >0$.
\begin{enumerate}

\item  $\ell < \ell _{ms(i)}$. Open surfaces are the only ones in existence,
No disks are possible while surfaces
with outer cusp exists (Fig. \ref{fig4}a1-2).

\item $\ell =\ell _{ms(i)}$. An infinitesimally thin,
unstable ring located at $r_{ms(i)}$ exists also open surfaces with
outer cusps remain in the picture (Fig. \ref{fig4}b1-2).

\item $\ell _{ms(i)}<\ell <\ell _{mb}$. Closed surfaces comes into existence.
Many equilibrium configuration even without any
cusp is now possible. We also have surfaces with inner and outer cusp (Fig. \ref{fig4}c1-2).

\item $\ell =\ell _{mb}$. Here also equilibrium configurations without cusps are possible.
An surface come into
existence with both inner and outer cusp. Also along with inflow into
black hole due to mechanical non-equilibrium,
an outflow from disk also occur. This comes due to the repulsive nature of the cosmological parameter.
This is an astrophysically important scenario (Fig. \ref{fig4}d1-2).

\item $\ell _{mb}<\ell <\ell _{ph(c)}$. Equilibrium configurations are
possible but accretion is not possible. Since we have no closed
surface with inner cusp. The surface with inner cusp is an open
equipotential surface, while that with outer cusp
is an closed equipotential surface. This makes outflow from the disk possible (Fig. \ref{fig4}e1-2).

\item $\ell =\ell _{ph(c)}$. The potential $W(r,\theta =\pi/2,y)$ diverges at the photon circular orbit.
Thus inner cusp completely disappears. Closed equipotential surfaces with outer
cusp still exists enabling outflow from
disk (Fig. \ref{fig4}f1-2).

\item $\ell =\ell _{ms(o)}$. An infinitesimally thin,
unstable ring located at $r_{ms(o)}$ exists with coalescing outer and central cusp (Fig. \ref{fig4}g1-2).

\item $\ell >\ell _{ms(o)}$. Only open equipotential surfaces exist with no cusp. (Fig. \ref{fig4}h1-2)

\end{enumerate}

\item $y=y_{e}.~$ For this special $y$ value the variation of $\ell _{K}^{2}$
and $\ell _{ph}^{2}$ has been illustrated in Fig. \ref{fig3}b.
For this case also we obtain all the hypersurfaces with similar nature. All of them falls into
a single class, given by,
\begin{enumerate}

\item $\ell =\ell _{ph(c)}=\ell _{ms(o)}$. Here there exists no inner cusp only an outer cusp mixing
with the center exists (Fig. \ref{fig4}i1-2).

\end{enumerate}

\item $y_{e}<y<y_{ms}$ This situation is illustrated by Fig. \ref{fig3}c. However
in this situation many other choices comes into existence. We classify them below,

\begin{enumerate}

\item $\ell _{mb}<\ell <\ell _{ms(o)}$. This condition is equivalent to the condition (1e)
illustrated above.

\item $\ell =\ell _{ms(o)}.$ This situation is slightly different. Here there exists an inner
cusp of an open equipotential surface, while the outer cusp gets mixed with center corresponding to a
thin ring at the radius $r_{ms(o)}$ (Fig. \ref{fig4}j1-2).

\item $\ell _{ms(o)}<\ell <\ell _{ph(c)}$. This condition has only open equipotential
surfaces among them one has an inner cusp (Fig. \ref{fig4}k1-2).

\end{enumerate}

\item$y_{ms}<y<y_{c}$. For this choice of cosmological parameter
the variations of $\ell _{K}^{2}$ and $\ell _{ph}^{2}$ are shown in Fig. \ref{fig3}d.
Note that in this situation $\ell _{K}^{2}$ has only a declining part. Thus we can
have only maxima of the potential and existence of open equipotential surfaces. This
leads to the fact that in these spacetimes stable circular geodesics do not exist. In this
case there is only two physically meaningful interval.

\begin{enumerate}

\item $\ell <\ell _{ph(c)}$. This situation has equipotential surfaces identical to
the situation illustrated in (1a).

\item $\ell \geq \ell _{ph(c)}$ This is similar to the situation (2a) discussed earlier.

\end{enumerate}
\end{enumerate}
We should also mention in this connection that the following
numerical values are obtained for angular momentum densities,
which are important for the study of fluid orbiting around black
hole. The angular momentum density for inner marginally stable
orbit is $\ell _{ms(i)}=3.03965$, while that for outer marginally
stable orbit is given by, $\ell _{ms(o)}=4.9$ for cosmological
parameter being $y=10^{-6}$. Two other important angular momentum
densities are that of marginally bound orbit with, $\ell
_{mb}=3.28852$ and for photon circular orbit as, $\ell
_{ph(c)}=4.27392$. These expressions are of quiet importance to study
the structure of accretion disk.

\subsection{Charged Black Hole in Dilaton Gravity}\label{equidilaton}

Equilibrium configurations originating from test particle fluid, which is rotating in a
given spacetime are determined by the equipotential surfaces, where the gravitational
and inertial forces are being compensated by the pressure gradient.
(For an axially symmetric spacetime the rotation axis of the equilibrium
configuration coincides with
the axis of symmetry for the spacetime, while for a spherically symmetric
case this axis can be any radial line; see for example \cite{Stuchlik00}).

The influence of a non-zero dilaton charge coming from an effective string theory on the character
of the equipotential surfaces of marginally stable orbits have been studied, for configuration
rotating around such black holes. Static uncharged black hole in general relativity is
described by well known and well studied Schwarzschild solution.
However even for large mass black hole (compared to
Planck mass) the Schwarzschild solution is a good approximation to describe
uncharged black hole in string theory except for regions near the singularity.
This situation is completely different for the Einstein-Maxwell solution
in string inspired theory due to dilaton
coupling.

The dilaton couples with $F^{2}$ with implication that every solution having non zero $F_{\mu \nu}$
couples with dilaton. Thus the charged black hole solution in general relativity
(the Reissner-Nordstr\"{o}m solution) appears in string theory with
presence of dilaton. Thus the effective four dimensional low energy action is
given by,
\begin{equation}\label{dil01}
S=\int d^{4}x \sqrt{-g}\left[-R+e^{-2\Phi}F^{2}+2(\nabla\Phi)^{2}\right]
\end{equation}
where $F_{\mu \nu} $ is the Maxwell field and we have set other gauge and
antisymmetric tensor field, e.g. the Kalb-Ramond field $H_{\mu \nu \rho}$ to zero in order
to focus on dilaton field $\Phi$ \cite{Garfinkle91,Coleman83,Vega88,Witten62,Bekenstein}.
Extremizing the above action with respect to the $U(1)$ potential $A_{\mu}$,
dilaton field $\Phi$ and metric $g_{\mu \nu}$ lead to the following field equations,
\begin{eqnarray}\label{eq15}
\nabla _{\mu} \left(e^{-2\Phi}F^{\mu \nu} \right)&=&0
\\
\nabla ^{2}\Phi +\frac{1}{2}e^{-2\Phi}F^{2}&=&0
\label{eq15a}
\\
R_{\mu \nu}=2\nabla _{\mu}\Phi \nabla _{\nu}\Phi &+&2e^{-2\Phi}F_{\mu \lambda}F^{\lambda}_{\nu}-\frac{1}{2}g_{\mu \nu}e^{-2\Phi}F^{2}
\end{eqnarray}
The static spherically symmetric solution to the above field equations yield the line element as
\cite{Garfinkle91}:
\begin{equation}\label{eq16}
ds^{2}=-(1-\frac{2M}{r})dt^{2}+\frac{1}{(1-\frac{2M}{r})}dr^{2}+r(r-e^{2\Phi_{0}}\frac{Q^{2}}{M})d\Omega^{2}
\end{equation}
where, $d\Omega^{2}=d\theta^{2}+sin^{2}\theta d\phi^{2}$. Due to isometry
we can confine our motion in the equatorial plane such that, $d\Omega^{2}=d\phi^{2}$
[i.e. $\theta =\pi/2$].
Here $\Phi_{0}$ gives the asymptotic value of dilaton field and $Q$ represents the black hole
charge. This solution is almost identical to the Schwarzschild metric with the difference
being that areas of two spheres depend on Q. The surface
$r=\frac{Q^{2}e^{2\Phi_{0}}}{M}$ is singular and $r=2M$ is the regular event horizon.
The solution of the scalar field $\Phi$ as a function of $r$ in terms of its asymptotic value $\Phi _{0}$ can be obtained from Eq. (\ref{eq15a}) leading to the following result \cite{Garfinkle91}:
\begin{equation}\label{eq17}
e^{-2\Phi}=e^{-2\Phi _{0}}-\frac{Q^{2}}{Mr}
\end{equation}
From the above result note that we have the following limit,
$r\rightarrow \infty$ leading to $\Phi \rightarrow \Phi _{0}$.
Thus the dilaton charge can be defined as,
\begin{equation}\label{dil02}
D=\frac{1}{4\pi}\int d^{2}\sigma^{\mu}\nabla_{\mu}\Phi
\end{equation}
where the integral is over the
two sphere located at spatial infinity and $\sigma^{\mu}$ is the
normal to the corresponding two sphere at spatial infinity.
For the charged black hole we are considering this leads to,
\begin{equation}\label{eq18}
D=-\frac{Q^{2}e^{2\Phi_{0}}}{2M}
\end{equation}
Note that dilaton charge D depends on the asymptotic value of dilaton field $\Phi _{0}$,
which is determined solely by M and Q
and is always negative. 

One point must be mentioned while passing, that this 
charge originates from coupling of electromagnetic field with a scalar field, 
the actual charge present in the system is the electromagnetic charge Q. If there 
were no electromagnetic charge the dilaton charge D defined through Eq. (\ref{eq18}) would 
have been zero. Hence this dilaton charge is not a scalar charge it's a coupling 
between the electromagnetic charge and the scalar field at asymptotic infinity. This 
charge is also responsible for long range attractive force between black holes 
\cite{Garfinkle91}.

Note that the actual dependence on dilaton field
depends on Planck scale described by, $e^{-\Phi /M_{pl}}$.
Since we have worked in the unit $M_{pl}\sim 1$ the
term modified to $e^{-\Phi}$. As $\Phi \rightarrow \Phi _{0}\sim M_{pl}$, this term
become significant. For notational convenience we shall define a quantity
$q=-D$ and by virtue of above discussion $q$ is always positive. Hence our solution is
parameterized by the variable $q$ and choosing the dimensionless expression such that
$M=1$ we get the line element as
\begin{equation}\label{eq19}
ds^{2}=-\left(1-\frac{2}{r}\right)dt^{2}+\left(1-\frac{2}{r} \right)^{-1}dr^{2}+
r\left(r-2q\right)d\Omega ^{2}
\end{equation}
Then we have two horizons in our system given by $r_{h1}=2$ and $r_{h2}=2q$. We shall
took the choice in which $r_{h1}>r_{h2}$ or equivalently $q\leq1$.
\begin{figure*}
\begin{center}
(a)
\includegraphics[height=2in, width=2.5in]{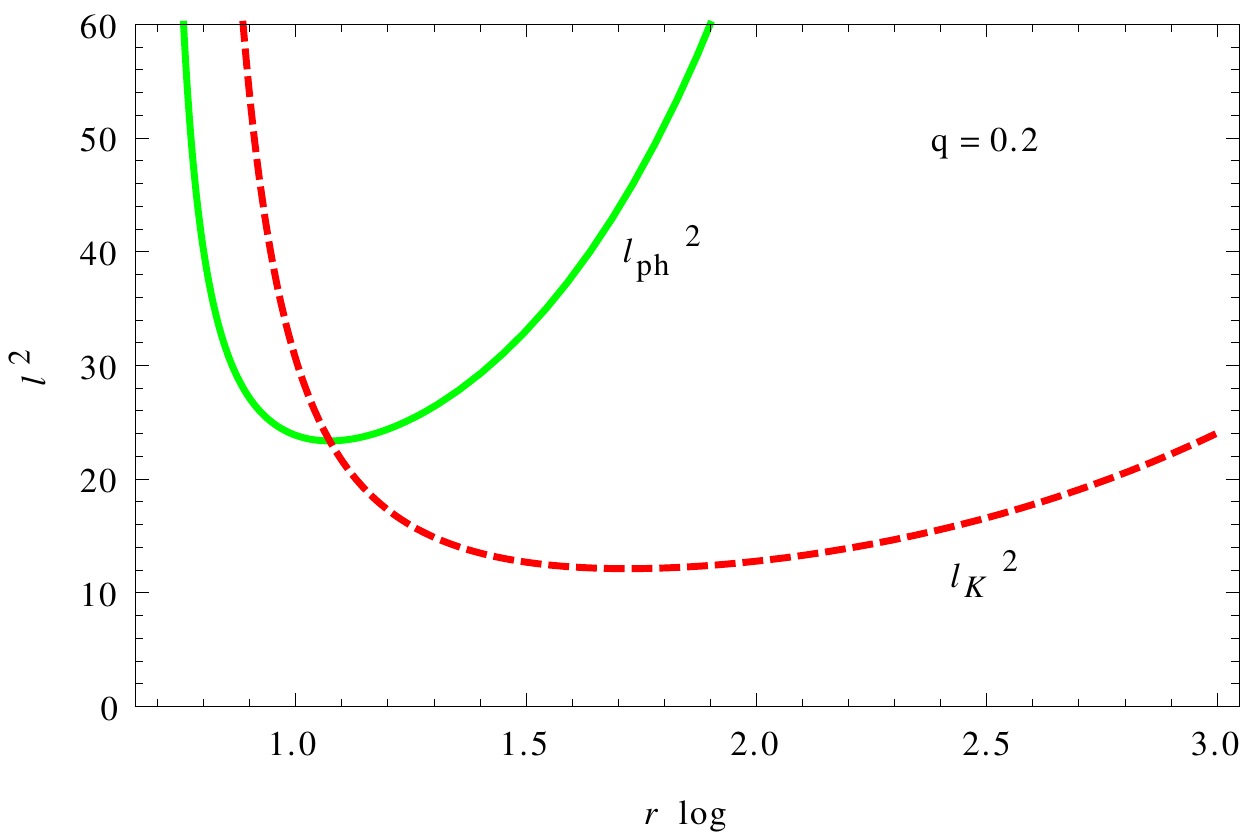}~~~
(b)
\includegraphics[height=2in, width=2.5in]{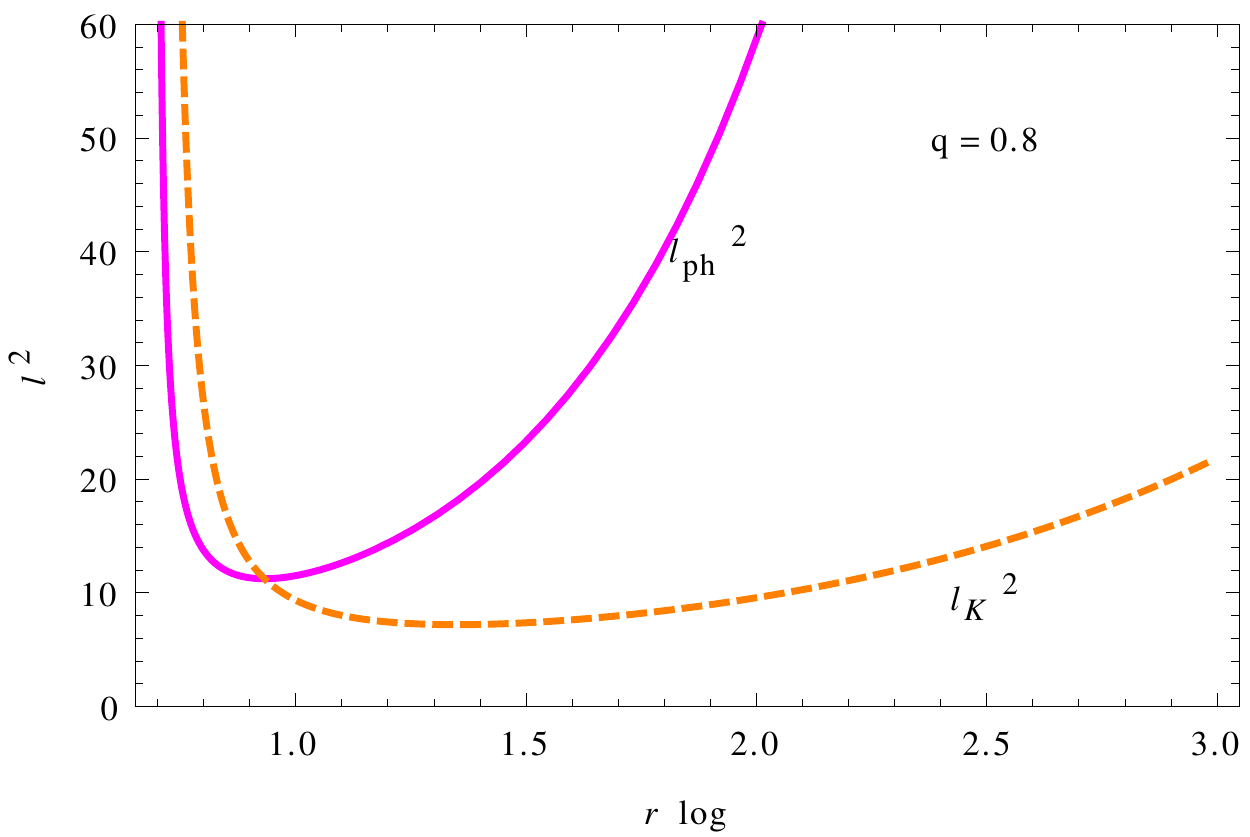}\\

\caption{The figures show the behavior of $l_{ph}^{2}(r,q)$ and $l_{K}^{2}(r,q)$
with radial variable for different values of dilaton charge $q$. All of them are
in the units of $M$. Here we have all the figures in the range $q<1$. The curve for
$\ell _{K}^{2}(r,y)$ has only growing part showing only existence of inner ring.
Thus no inner cusp exists in these spacetime.
\label{fig5} }

\end{center}
\end{figure*}
Thus equilibrium configurations are possible only for $r>2$. Now the equipotential
surfaces are possible only in these spacetimes. Now, the equipotential surfaces are
determined by the result,
\begin{equation}\label{eq20}
W(r,\theta)=\frac{1}{2}\ln \frac{\left[r(r-2)(r-2q)\sin^{2}\theta\right]}{\left[r^{2}(r-2q)\sin^{2}\theta-(r-2)\ell ^{2}\right]}
\end{equation}
and
\begin{equation}\label{eq21}
\frac{d\theta}{dr}=\frac{r^{2}(r-2q)^{2}\sin^{2}\theta -\ell ^{2}(r-2)^{2}(r-q)}{\ell ^{2}r(r-2)^{2}(r-2q)}\tan\theta
\end{equation}
For $q=0$ the above results reduces to the well known
Schwarzschild result \cite{Jaroszynski80}. The best insight
into the nature of the $\ell =\textrm{constant}$ fluid
configurations are obtained by examining the behavior of
$W(r,\theta)$ in the equatorial plane ($\theta =\pi /2$). There
are two reality conditions on $W(r,\theta =\pi /2)$:
\begin{equation}\label{eq22}
r>2; r>2q
\end{equation}
and
\begin{equation}\label{eq23}
r^{2}(r-2q)>r^{2}\ell ^{2}
\end{equation}
The first condition is just the statement that the fluid should remain outside the black
hole horizons and we shall calculate it's property in that non singular region of spacetime.
While the second condition implies,
\begin{equation}\label{eq24}
\ell ^{2}\leq \ell ^{2}_{ph}(r;q)\equiv \frac{r^{2}(r-2q)}{r^{2}}
\end{equation}
This function $\ell _{ph}^{2}(r;q)$ can be thought of as an
effective potential of the photon geodesic motion, also note that
$\ell = \frac{U_{\phi}}{U_{t}}$ has a close  correspondence with
the impact parameter for photon geodesic motion \cite{Stuchlik99}. 
Further the condition to obtain local extrema
of the potential $W(r,\theta =\pi /2)$ is identical to the
condition for vanishing of pressure gradient ($\partial
U_{t}/\partial r=0, \partial U_{t}/\partial \theta =0$). Since at
the equatorial plane we have $\partial U_{t}/\partial \theta =0$
independently along with the criteria $\ell =\textrm{constant}$,
and the following relation:
\begin{equation}\label{eq25}
\frac{\partial U_{t}}{\partial r}=\frac{r^{2}(r-2q)^{2}-\ell ^{2}(r-2)^{2}(r-q)}{\left[r^{2}(r-2q)-\ell ^{2}(r-2)\right]^{3/2}\left[r(r-2)(r-2q)\right]^{1/2}}
\end{equation}
\begin{figure*}
\begin{center}

(a1)
\includegraphics[height=2in, width=2.5in]{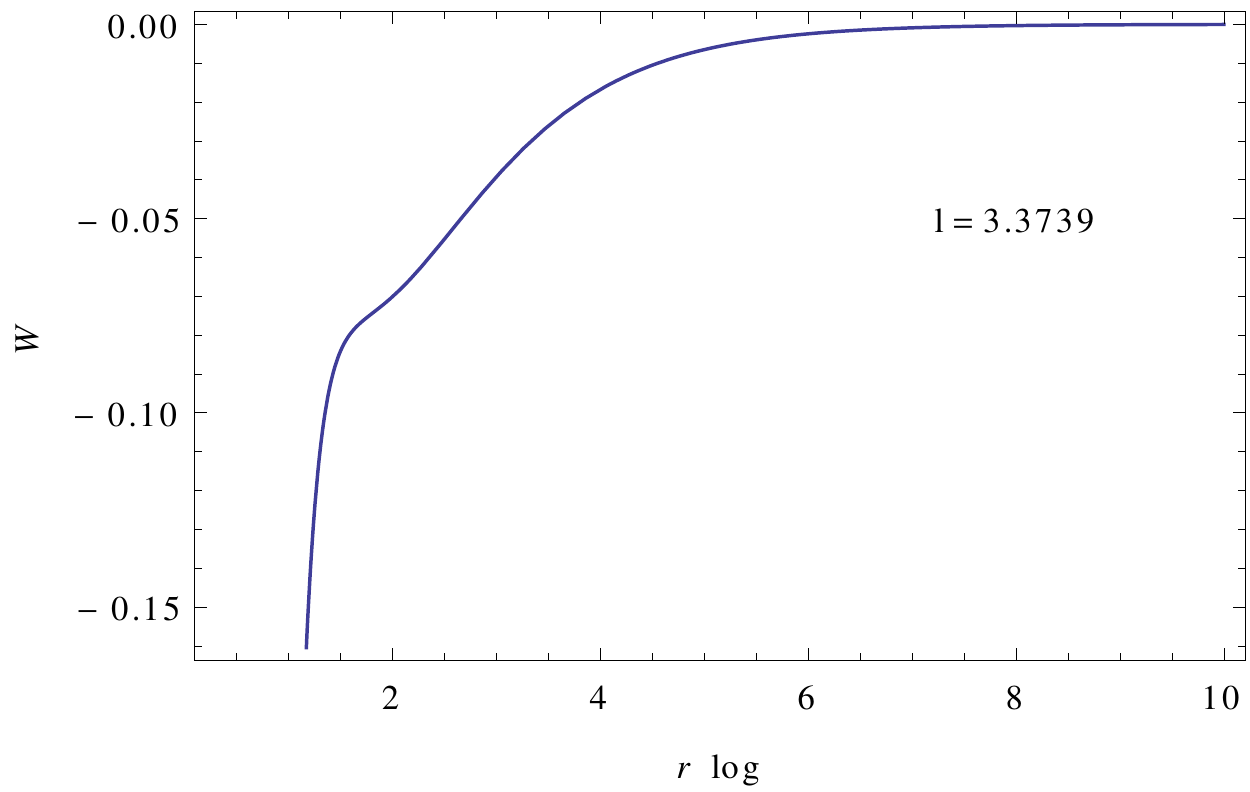}~~
(a2)
\includegraphics[height=2in, width=2.5in]{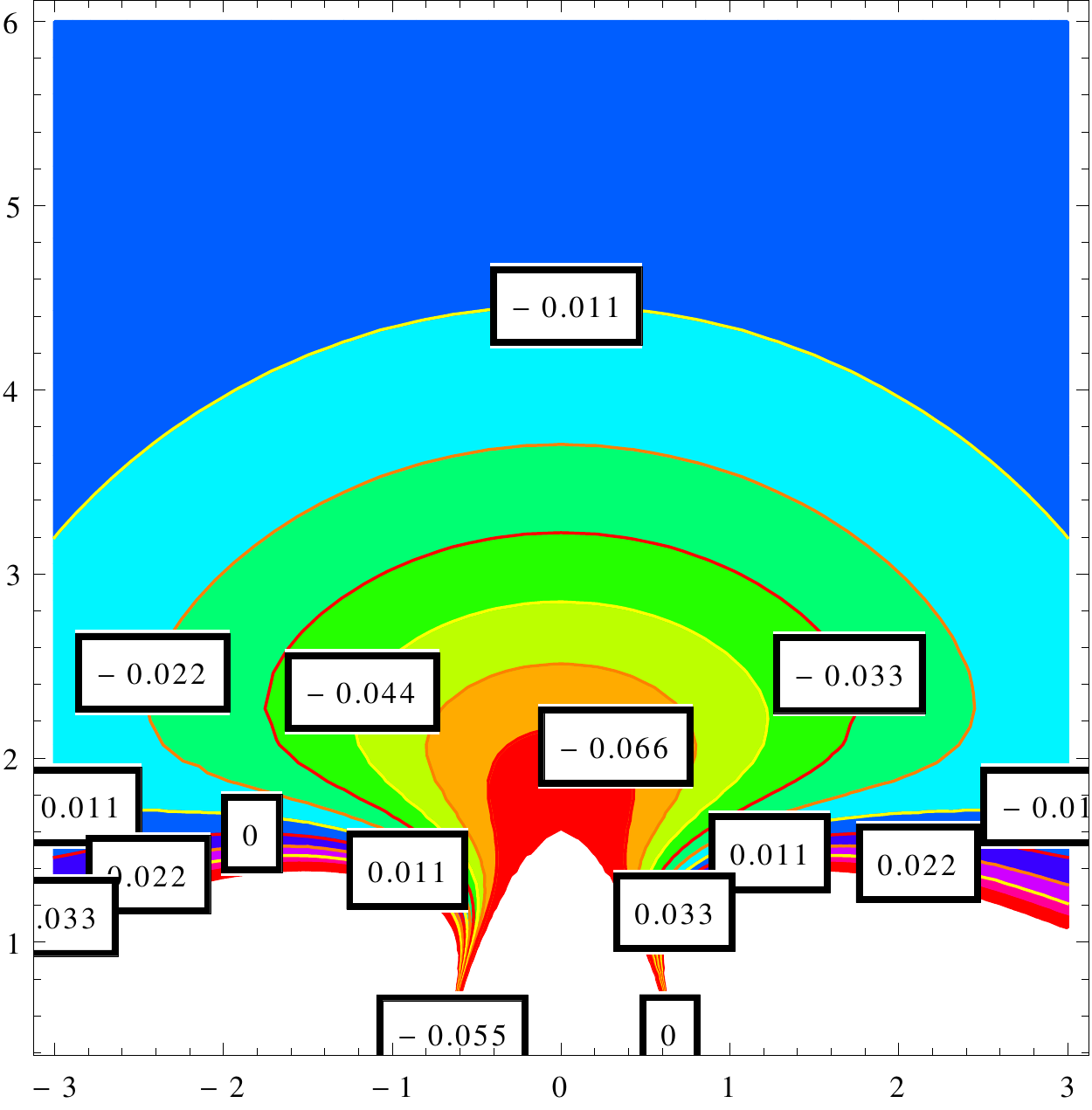}\\
(b1)
\includegraphics[height=2in, width=2.5in]{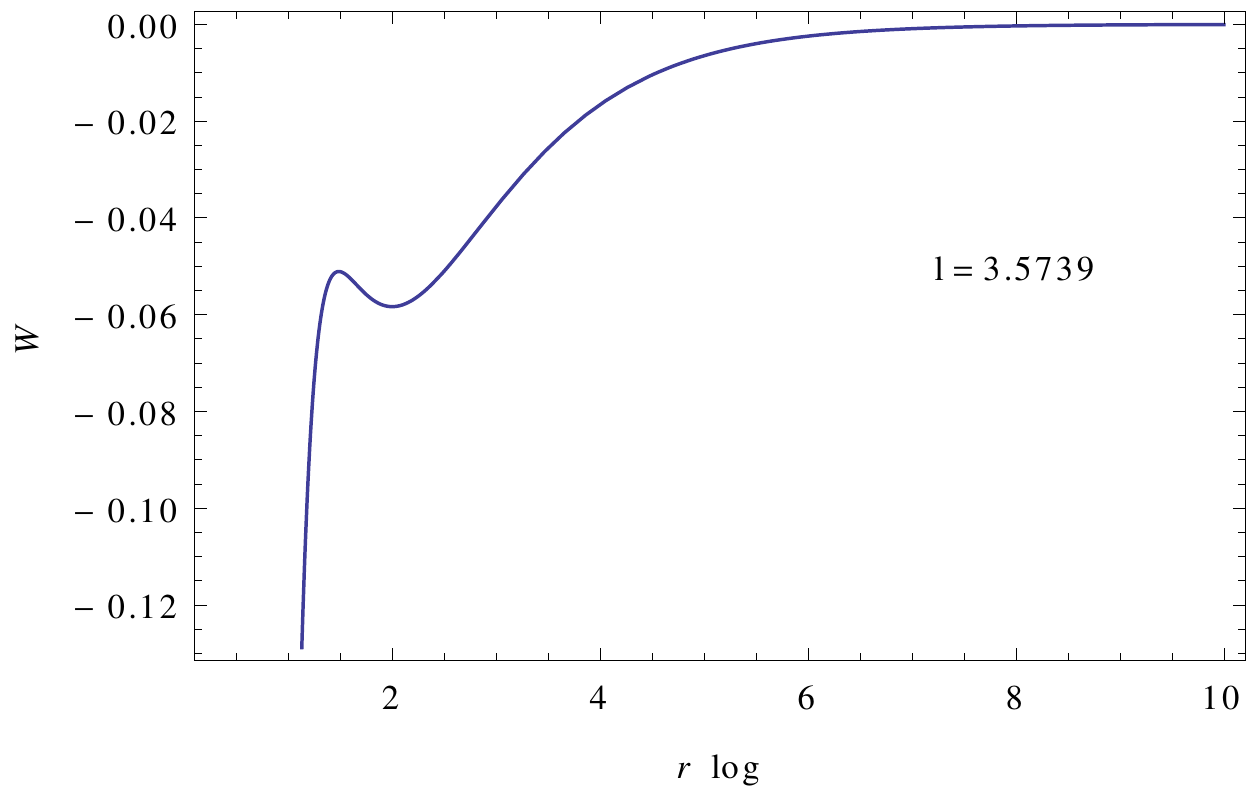}~~
(b2)
\includegraphics[height=2in, width=2.5in]{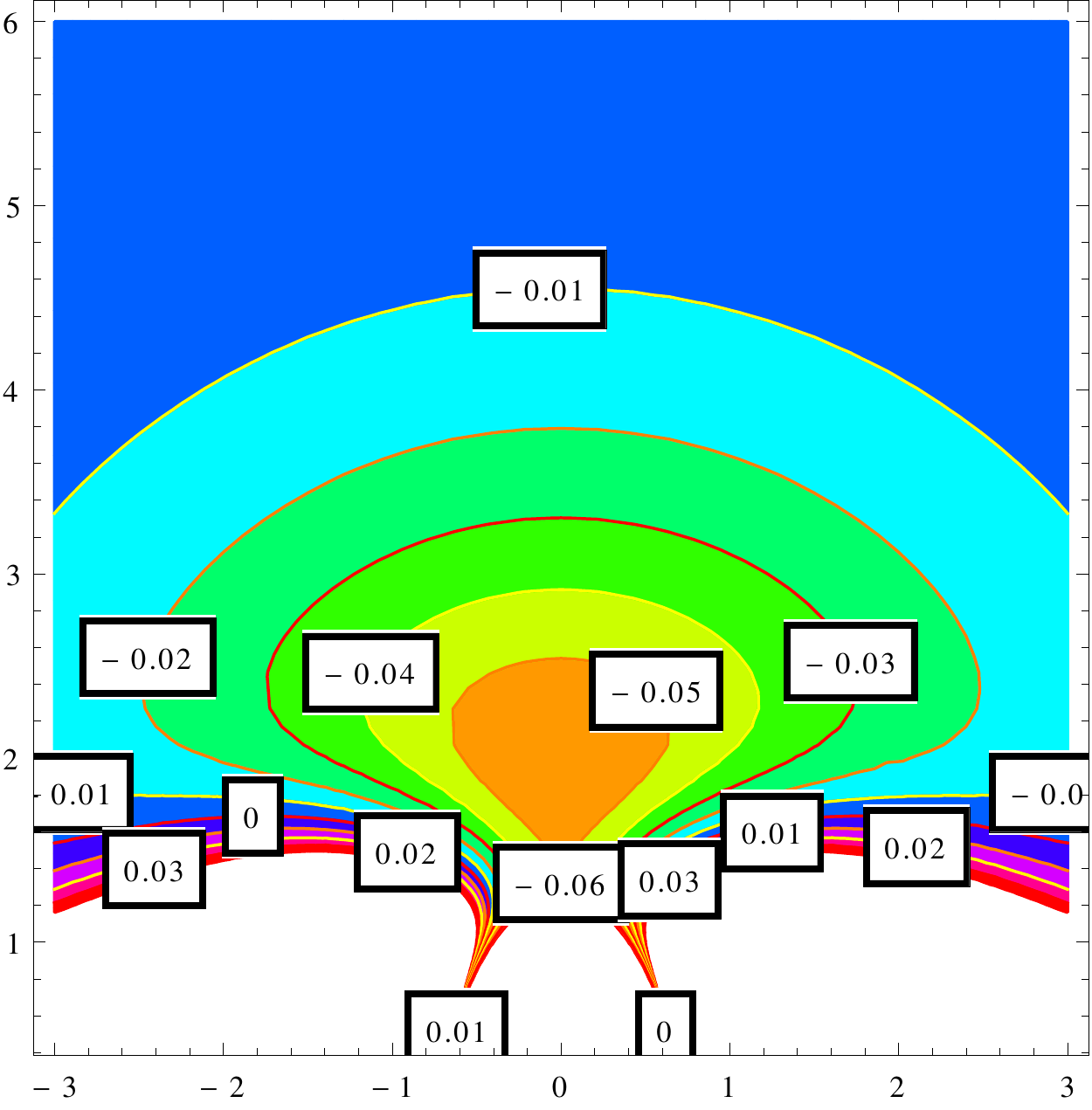}\\
(c1)
\includegraphics[height=2in, width=2.5in]{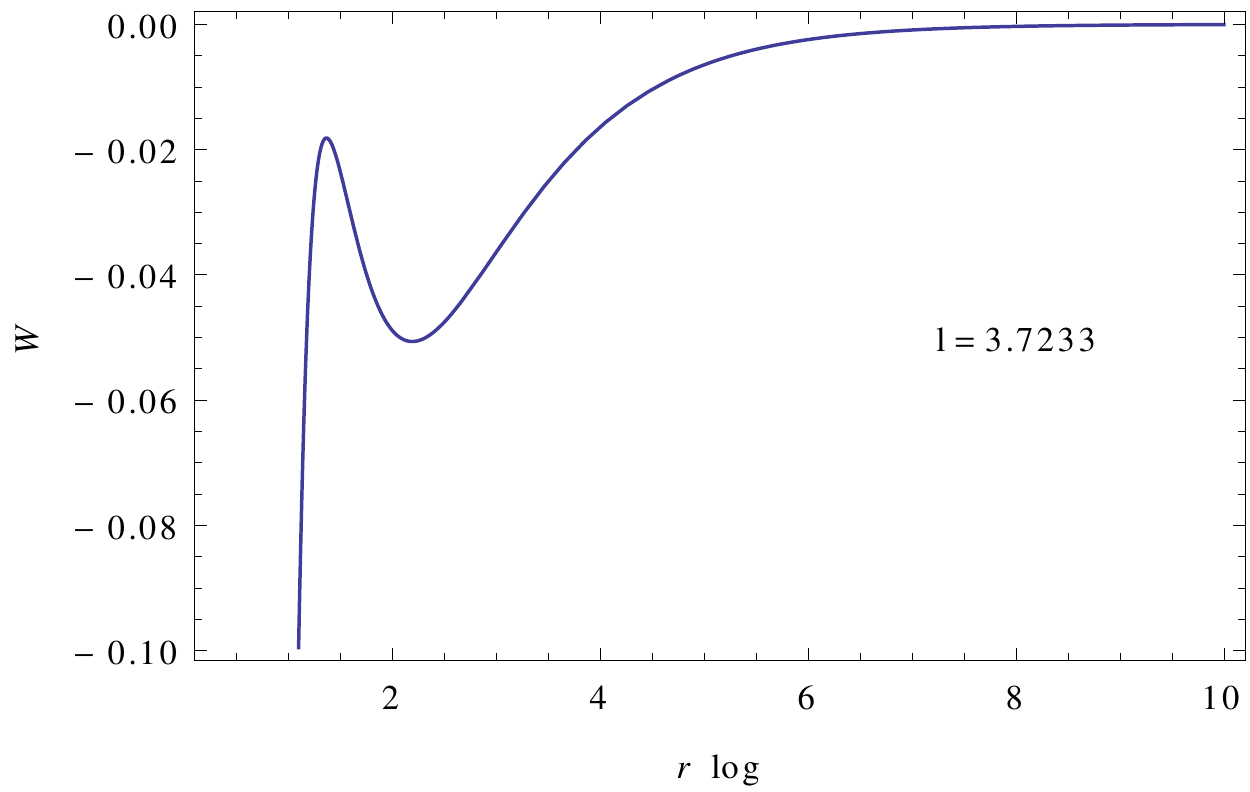}~~
(c2)
\includegraphics[height=2in, width=2.5in]{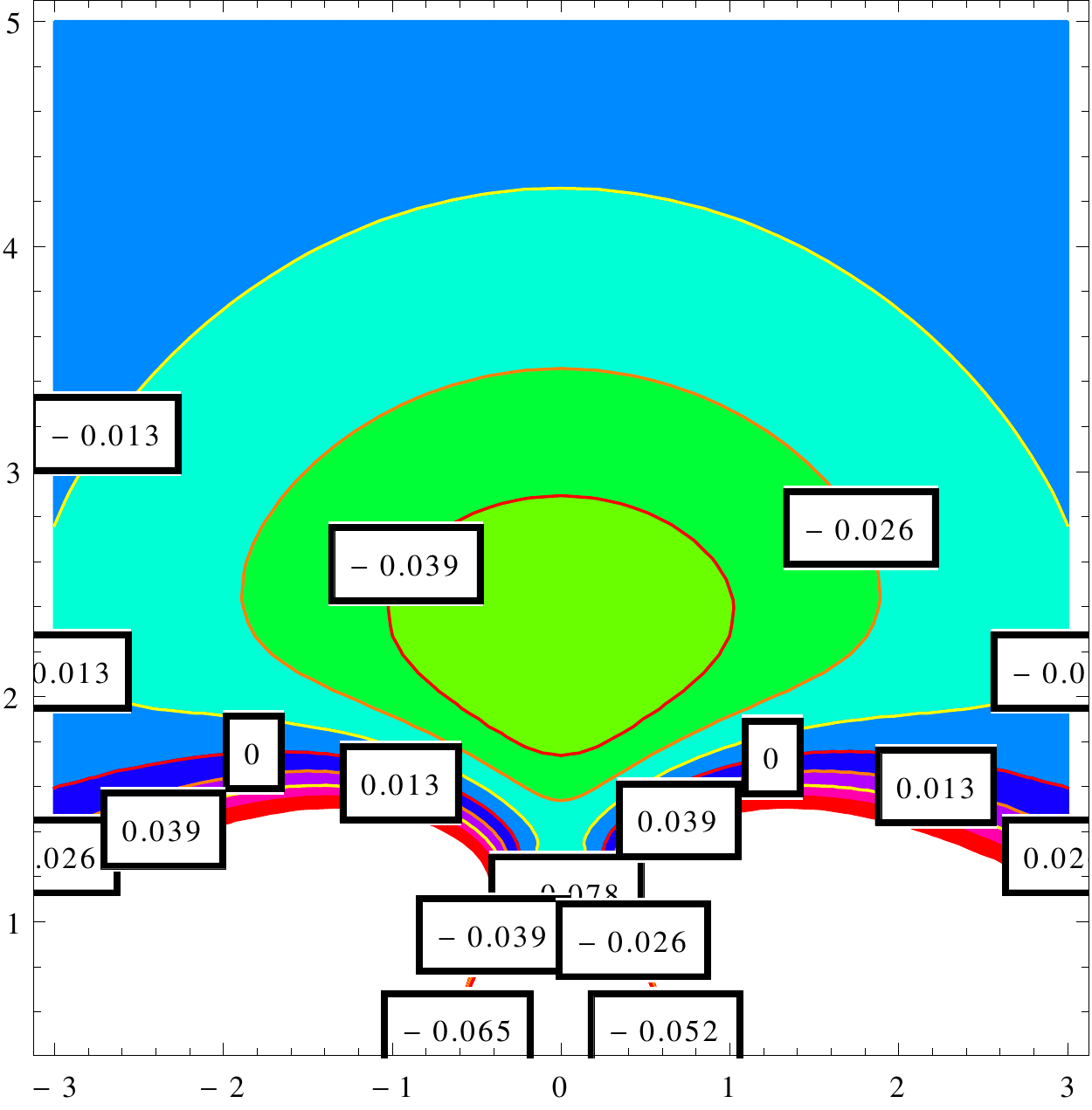}\\
(d1)
\includegraphics[height=2in, width=2.5in]{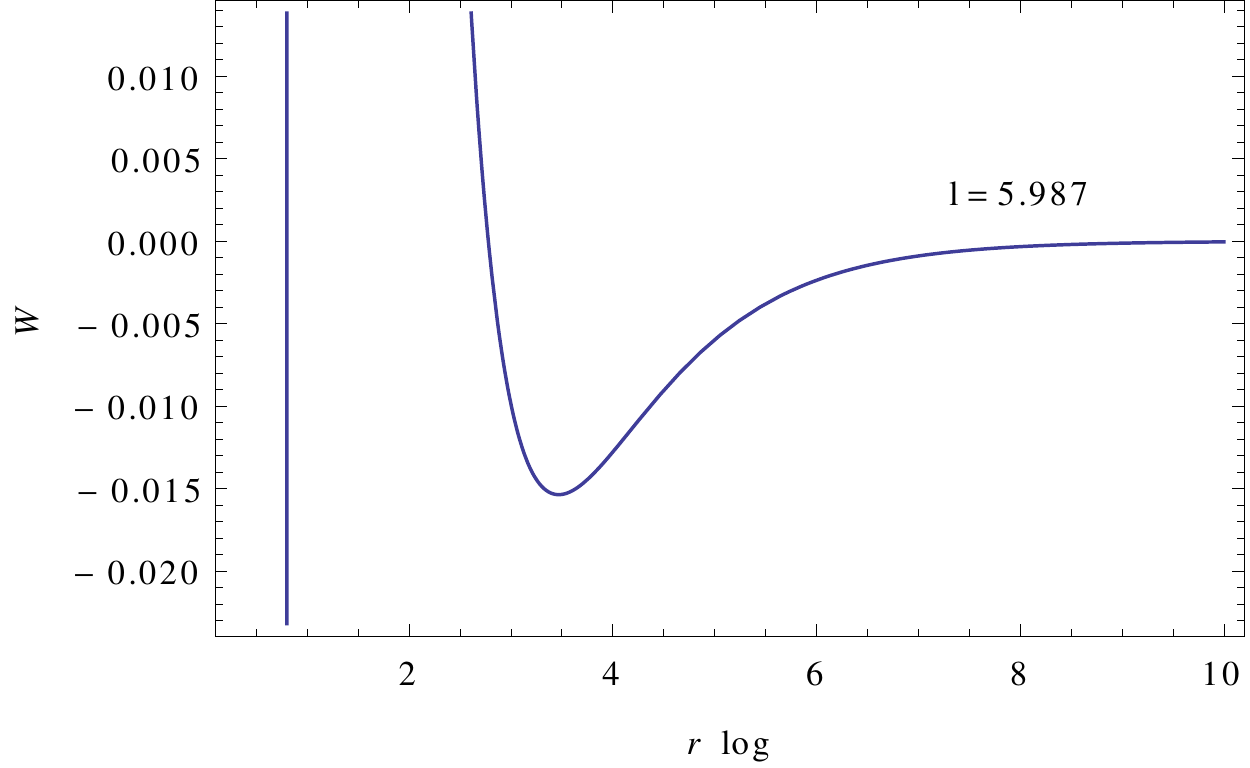}~~
(d2)
\includegraphics[height=2in, width=2.5in]{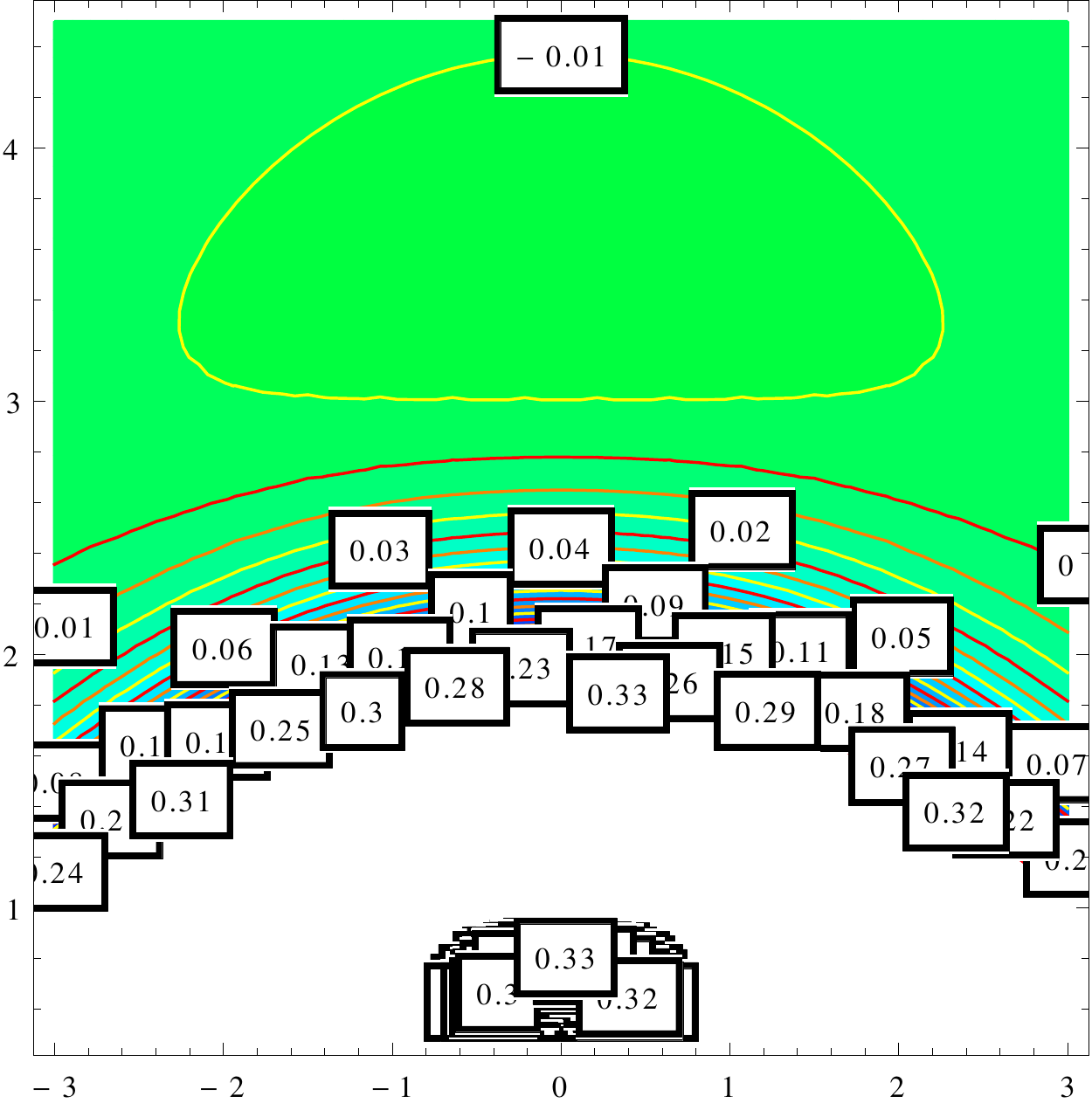}\\

\caption{The figures (a)-(d) show the variation of potential
$W(r,\theta =\pi/2,q)$ with $log~r$ in the `1' part and the contour plot of $W$ with $log(r~\sin \theta)$
and $log(r~\cos \theta)$ in the `2' part. Each figures are for values of dilaton charge $q<1$. Here only
central rings exists, while no inner cusp comes into existence. \label{fig6} }

\end{center}
\end{figure*}
From which we arrive at the particular expression for angular momentum density,
\begin{equation}\label{eq26}
\ell ^{2}=\ell _{K}^{2}(r;q)\equiv \frac{r^{2}(r-2q)^{2}}{(r-q)(r-2)^{2}}
\end{equation}
The extrema for $W(r,\theta =\pi /2)$ correspond to the spacetime points, where fluid moves
along a circular geodesics, since $\ell _{K}^{2}(r;q)$ relates to the distribution
of the angular momentum density of circular geodesic orbits. Clearly this leads to,
\begin{equation}\label{eq27}
W_{extr}(r,\theta =\pi /2;q)=\ln E_{c}(r,y)
\end{equation}
where
\begin{equation}\label{eq28}
E_{c}(r,q)=\frac{(r-2)\sqrt{(r-q)}}{\sqrt{r}\sqrt{(r-q)(r-2)-(r-2q)}}
\end{equation}
is defined as the specific energy along these circular geodesics.
Important properties of the potential $W(r,\theta)$ are determined
by its behavior at the equatorial plane, and, specially by the
properties of the functions $\ell _{ph}^{2}(r;q)$ and $\ell
_{K}^{2}(r,q)$. Discussion of these properties enable us to
classify the Dilaton gravity spacetime according to the properties
of equipotential surfaces of test particle fluid. For pure
Schwarzschild spacetime ($q=0$) the analysis can be found in Ref.
\cite{Kozlowski78}.

From the analytical form of $\ell _{ph}^{2}(r,q)$ it is evident that the
quantity diverges at the black hole horizon ($r=2$). However it shows no
such divergence for the inner horizon given by $r=2q$. The local minimum of
the function $\ell _{ph}^{2}(r,q)$ corresponds to the radius of photon circular
orbit and has the following expression,
\begin{equation}\label{eq29}
r_{ph}=\frac{3+q\pm \sqrt{(q+3)^{2}-16q}}{2}
\end{equation}
with the impact parameter,
\begin{equation}\label{eq30}
\ell _{ph(c)}^{2}=\ell _{ph(min)}^{2}\equiv r^{2}(r-q)
\end{equation}
The function $\ell _{K}^{2}(r,q)$, determining the Keplerian (geodesic) circular orbits,
has a zero point at the so called static radius given by,
\begin{equation}\label{eq31}
r_{s}=2q
\end{equation}
which coincides with the inner horizon, also note that the function 
has divergent nature at $r=2$ and as well as at $r=q$.
The function $\ell _{K}^{2}(r;q)$ diverges at the black hole 
horizon i.e. $\ell _{K}^{2}(r\rightarrow 2)\rightarrow +\infty$. Since,
\begin{eqnarray}\label{eq32}
\frac{\partial \ell _{K}^{2}}{\partial r}&=&\frac{4r(r-2q)}{(r-2)^{2}}
\nonumber
\\
&-&\frac{r^{2}(r-2q)^{2}(3r-2-2q)}{(r-q)^{2}(r-2)^{3}}
\end{eqnarray}
the local extrema of $\ell _{K}^{2}(r,q)$ are given by the condition, $q_{ms}=1$
determining the marginally stable circular orbits. For $q<q_{ms}$, there exists an inner (outer)
marginally stable circular geodesic at $r_{ms(i)}$ ($r_{ms(o)}$).

Other special values of $q$ corresponds to the situation where the value of the minimum of
$\ell _{ph}^{2}(r,q)$ equals the maximum of $\ell _{K}^{2}(r,q)$. We denote this value by
$q_{e}$ and is a solution of the algebraic equation,
\begin{equation}\label{eq34}
q_{e}^{5}-22q_{e}^{4}+80q_{e}^{3}-64q_{e}^{2}-32=0
\end{equation}
However this situation is less important as the above equation has no physical
solution. Thus we will not have any inner cusp present in the system. Hence this
has less attractive features, still interesting which we will present now.

In this situation the behavior of $\ell _{ph}^{2}$ and $\ell
_{K}^{2}$ are similar to that of RN scenario. while we have only
four available situations. As illustrated in Fig. \ref{fig5} we
have a minimum for $\ell _{K}^{2}$ having value, $\ell
_{ms}=3.5739$. We also have a minimum in $\ell _{ph}^{2}$ which
corresponds to the value, $\ell _{ph(c)}=4.2786$. Thus all
together we have four cases in this dilaton gravity model from
behavior of the potential $W(r,\theta =\pi/2,q)$. These cases are
given for following intervals of $\ell$:
\begin{enumerate}

\item $\ell < \ell _{ms}$. We have only open equipotential surfaces in this angular
momentum range [see Fig. \ref{fig6}a1-2].

\item $\ell =\ell _{ms}$. An infinitesimally thin and unstable rings comes to existence at
the marginally stable radius [see Fig. \ref{fig6}b1-2].

\item $\ell _{ms}<\ell \leq \ell _{ph(c)}$. Closed equipotential surfaces comes
into existence, with one such surface having cusp, allowing inflow to the black hole
[see Fig. \ref{fig6}c1-2]

\item $\ell > \ell _{ph(c)}$. Closed equipotential surfaces exist, however cusps does not exist. Thus
in the near horizon region equipotential surfaces cannot cross equatorial plane
[see Fig. \ref{fig6}d1-2].

\end{enumerate}
Thus we have obtained structure of equipotential surfaces in
dilaton gravity. How the inflow occur in this spacetime along with
necessary criteria on the angular momentum density. Also behavior
in determining the angular momentum density which admits no
minima. Hence this can be thought of having less structure than
the previous one still having quiet interesting features.

\subsection{Einstein-Maxwell-Gauss-Bonnet Gravity}\label{FluidEMGB}

In high energy physics, assuming spacetime to have more than four
dimensions is a common practice. In these scenarios, the spacetime
is assumed to be a four dimensional brane, embedded on a higher
dimensional bulk. Though ordinary matter fields are confined to
these branes, gravity can interact via bulk as well. The EH action
needs to be modified as well and the most natural choice would be
to include higher order terms in the action. A famous second order
term is called the Gauss Bonnet term, under whose presence the
modified action looks like,
\begin{eqnarray}
S&=&\int dx^{5}\sqrt{-g}\Big[R+\alpha \Big(R_{abcd}R^{abcd}-4R_{ab}R^{ab}
+R^{2}\Big)+F_{ab}F^{ab} \Big]
\end{eqnarray}
where $R$, $R_{ab}$ and $R_{abcd}$ are Ricci scalar, Ricci tensor and
Riemann tensor respectively. $F_{ab}$ is the electromagnetic tensor field and
$\alpha$ being the GB coupling coefficient with dimension of length squared.
In order to get equations of motion we need to vary the action with respect
to the metric $g_{ab}$ and electromagnetic field $F_{ab}$ respectively.
This leads to \cite{Dehghani04},
\begin{eqnarray}
R_{ab}&-&\frac{1}{2}g_{ab}R-\alpha \Big[\frac{1}{2}g_{ab}\Big(R_{pqrs}R^{pqrs}
\nonumber
\\
&-&4R_{mn}R^{mn}+R^{2}\Big)-2RR_{ab}+4R_{am}R^{m}_{b}
\nonumber
\\
&+&4R^{mn}R_{ambn}-2R_{a}^{~pqr}R_{bpqr}\Big]=T_{ab}
\\
\nabla _{a}F^{ab}&=&0
\end{eqnarray}
where $T_{ab}$ is the usual stress tensor for electromagnetic field. The important thing to notice
is that the field equation only contains second order derivatives of the metric, no higher derivatives
are present. This is expected since Gauss-Bonnet gravity is a subclass of Lovelock gravity, which does not contain higher derivative terms of the Riemann tensor.

Now it is possible to obtain static spherically symmetric solution to this field equation having
the form of Eq. (\ref{geneq1}). It turns out that those solutions are
asymptotically de Sitter or anti-de Sitter.
Thus this situation would be identical to that in Sec. \ref{fluidfr}. This
immediately tells us that only a correspondence between the parameters in
this theory have to match with that of charged $F(R)$ scenario
discussed in an earlier section.

To this end we must mention that the solution obtained 
above corresponds to a five-dimensional spacetime, i.e. the line element takes the form: $ds^{2}=-f(r)dt^{2}+f^{-1}(r)dr^{2}+r^{2}d\Omega _{3}^{2}$, where $d\Omega _{3}^{2}=d\theta _{1}^{2}+\sin ^{2}\theta _{1}(d\theta _{2}^{2}+\sin ^{2}\theta _{2}d\theta _{3}^{2})$. However as we have been emphasizing 
that this solution can be interpreted from a brane world point of view, such that 
the visible brane is characterized by $\theta _{3}=\textrm{constant}$ hypersurface. 
Hence the solution reduces to four dimension with $(t,r,\theta _{1},\theta _{2})$ as a set of coordinates. 
Thus on the visible brane the spherically symmetric solution can be treated 
at the same footing to those discussed earlier.
Such spherically symmetric solutions are obtained in Ref. \cite{Dehghani04} and has the particular form
with reference to Eq. (\ref{geneq1}) as
\begin{equation}
f(r)=K+\frac{r^{2}}{4\alpha}\left[1\pm \sqrt{1+\frac{8\alpha \left(m+2\alpha \mid K \mid \right)}{r^{4}}-\frac{8\alpha q^{2}}{3r^{6}}} \right]
\end{equation}
where $K$ determines the scalar curvature of the spacetime, to which
we attribute a positive value. Then form solar system tests and neutrino
oscillation experiments \cite{Chakraborty14a,Chakraborty14b} 
it turns out that $\alpha^{-1}$ has stringent bounds.
Also in all astrophysical scenarios the distance from black hole $r$ is quiet large, thus for our study we can make a power series expansion of the terms inside the square root and arrive at the following form:
\begin{equation}
f(r)=1+\frac{r^{2}}{2\alpha}+\frac{m+2\alpha}{r^{2}}-\frac{q^{2}}{3r^{4}}
\end{equation}
Then we readily identify this with Eq. (\ref{FRC05}) leading to
the following mapping, $\alpha =-1/(2y)$. The possibility that
$\alpha$ may be negative has been discussed in detail by
\cite{Dehghani04b} and \cite{Corradini04}. With this
identification we can run all our machinery described in previous sections and obtain all the relevant physical quantities. 

All the numerical values for cosmological parameter discussed earlier is directly mapped to that of $\alpha$. We can also conclude that it has identical structure of equipotential surfaces, with existence of inner and outer cusp. Thus matter outflow can also occur in this topological dS or AdS black holes along with possibility of inflow to it.

Some numerical estimates can be made for this model following previous discussions. The angular momentum density for inner and outer marginally stable orbits are, $3.04$ and $4.9$ respectively with $\alpha=5\times 10^{5}$. Also marginally bound and photon circular orbits have similar expressions.

Thus for various alternative gravity theories we have obtained the equipotential surfaces for rotating perfect fluid. For some of them we have observed dynamical accretion disk requiring inflow, outflow and cusps. While in some cases the structure was simpler and do not have such dynamical behaviour. Thus we may conclude that black holes in different alternative theories affect accretion disk structure differently.

\section{Conclusions}

The effect of additional higher curvature correction terms in the EH action on the structure of equipotential surfaces for perfect fluid rotating around black holes in these modified gravity theories have been investigated. For that purpose we have performed the whole analysis for an arbitrary static spherically symmetric spacetime. Then we have applied the results obtained to different classes of alternative theories, with higher curvature terms. These theories include: charged $F(R)$ theory, dilaton induced gravity theory and finally the Einstein-Maxwell-Gauss-Bonnet gravity. In the $F(R)$ theory and EMGB theory the black hole solutions asymptotically behave as dS or AdS. The equilibrium structure of perfect fluid orbiting around these black holes in the above mentioned alternative gravity theories lead to modifications of the equipotential contours. Having provided the basic features let us now summarize the results:
\begin{itemize}

\item For the situation, $\ell =0$, we always have an open equipotential surface. The potential is being determined by the relation, $W=\ln \sqrt{f(r)}$. Thus different spherically symmetric solution leads to different equipotential surfaces.

\item Existence of outer cusp does not facilitate accretion onto the black hole. We need to have inner cusp for the equipotential surfaces as well, in order to have inflow to the black hole. Note that outer cusp is also necessary for accretion to occur.

\item Closed equipotential surfaces are possible if and only if the angular momentum lies in the range, $\ell \in \left(\ell _{ms(i)},\ell _{ms(o)} \right)$. The quantities, $\ell _{ms(i)}$ and $\ell _{ms(o)}$ represents the local minima and maxima of the angular momentum density respectively. For the charged $F(R)$ theory closed surfaces comes into existence provided the cosmological parameter $y$ satisfies, $y<y_{ms}=0.000692$. While for the EMGB gravity we obtain, $\alpha _{ms}=-782.47$. Note that these closed surfaces are necessary for existence of toroidal accretion disc.

\item The accretion by Paczy\'nski mechanism becomes possible, if angular momentum density lies in the range, $\ell \in \left(\ell _{ms(i)},\ell _{mb} \right)$, where $\ell _{mb}$ being the corresponding value of angular momentum density for marginally bound circular geodesics. In this scenario, the outflow from accretion disc becomes possible through both the inner and outer cusps. The existence of inner cusp leads to accretion flow directed towards the black hole.

\item For marginally stable configurations, due to overfilling of marginally closed equipotential surfaces efficient inflow and outflow occurs. However very near the horizon, the behaviour may change significantly.

\item Another physical situation of interest corresponds to the angular momentum range: $\ell \in \left(\ell _{mb},\ell _{ms(o)} \right)$. In this situation the flow down to the black hole becomes non-existent since open self-crossing surfaces come into existence.

\item Finally, for the situation where, $\ell >\ell _{ms(o)}$, the toroidal structure cannot exist. Equipotential surfaces exist however they are always open. Also the surfaces becomes narrower as it approaches the static radius which may have a significant effect on collimation of jets from the black hole.

\end{itemize}

These effects were most prominent for accretion of perfect fluid to the black holes in $F(R)$ theory and the EMGB theory. Both of which are asymptotically de Sitter for certain choice of parameter space. Richness in the structure of black hole accretion disk for asymptotically de Sitter solutions in GR were pointed out earlier in \cite{Stuchlik00}. Thus it is evident from this work that their generalization to alternative theories also come up with varied structure of the accretion disk. There are also asymptotically flat solutions like the dilaton black hole, which have new features compared to the GR scenario. However in this spacetime accretion disk does not posses all the structures which were present in the other two cases.  

Thus we observe that in Einstein gravity as well as in alternative theories spacetime with analogue of repulsive cosmological constant (for instance, $y$ in $F(R)$ model, $1/\alpha$ in EMGB theory) shows wide variety of phenomenon as the accretion disk structure is considered. Though the structure of the accretion disk changes considerably due to presence of higher curvature correction terms in the EH action it is very difficult to detect their signature astrophysically. This is mainly due to the reason that the parameter space of these alternative theories for being of astrophysical interest is quiet large compared to present day estimates of those parameters. For example, the black holes fuelling active galactic nuclei have typical masses in the range, $\sim 10^{8}M_{\odot}-10^{9}M_{\odot}$ which leads to an estimation of the cosmological parameter as, $y<10^{-10}$, which is very small and thus will not result in significant changes in the accretion disk structure. 

However the primordial black holes have a much larger value of the cosmological parameter. Thus accretion structure of them could be best hope to test these alternative theories. However this is beyond the scope of present day observations. Even though the departure of the equipotential contours in alternative theories from their general relativistic counterpart due to presence of higher curvature correction terms is an important effect but is unlikely to be observed in near future. 

To summarize, in this work we have discussed various possible candidates for alternative theories and the accretion disk structure in them. We have also left some questions unanswered. Among them the runaway instability criteria for toroidal accretion disks, i.e. stability with time has not been addressed as well as the self-gravitation influence of the disk is also not considered. These we leave for future.

\section*{Acknowledgement}

The author is funded by a SPM fellowship from CSIR, Govt. of India. The author would also like to thank Prof. T. Padmanabhan, Prof. Subenoy Chakraborty and Suprit Singh for helpful discussions. He also likes to thank the referees for providing valuable comments that have helped to improve the manuscript.


\end{document}